\documentclass[traditabstract,fleqn]{aa}
\usepackage{graphicx}
\usepackage{amssymb}
\usepackage{amsmath}
\usepackage{natbib}
\usepackage{subfigure}
\usepackage{txfonts}
\bibpunct{(}{)}{;}{a}{}{,} 

\begin{document}

\title{Structure of the nucleus of 1928+738}

\author{J. Roland\inst{1}, S. Britzen\inst{2}, E. Kun\inst{3,}\inst{4}, G. Henri\inst{5}, S. Lambert\inst{6} and A. Zensus\inst{2}  }

\authorrunning{Roland \& al.}
\titlerunning{Structure of the nucleus of 1928+738}

\institute{Institut d'Astrophysique, UPMC Univ Paris 06,
           CNRS, UMR 7095, 98\,bis Bd Arago, 75014 Paris, France
					\and
					Max-Planck-Institut f\"ur Radioastronomie,
           Auf dem H\"ugel 69, Bonn 53121, Germany
          \and
          Department of Experimental Physics, University of Szeged, 
            D\'om t\'er 9, H-6720 Szeged, Hungary
					\and
					Department of Theoretical Physics, University of Szeged, 
					 Tisza Lajos krt 84-86, H-6720 Szeged, Hungary
					\and
					Laboratoire d'Astrophysique, Observatoire de Grenoble, 
          414, Rue de la Piscine, Domaine Universitaire, 38400 Saint-Martin d'Hères, France
					\and
					Observatoire de Paris, SYRTE, CNRS, UPMC, CRGS, Paris, France
	        }

\offprints{J. Roland, \email{roland@iap.fr}}

\date{Received <> / Accepted <>}

\abstract{Modeling of the trajectories of VLBI components ejected by the nucleus of 
1928+738 shows the VLBI jet contains three families of trajectories, i.e. VLBI components 
are ejected from three different origins. The fit of components C1, C6 and C8 indicates that 
the nucleus of 1928+738 contains two binary black hole systems. 
The first binary black hole system is associated with the stationary components 
Cg and CS and is characterized by a radius $R_{bin,1} \approx 0.220$ mas; both black holes 
ejected VLBI components quasi regularly between 1990 and 2010. 
The second binary black hole system is not associated with stationary components 
and is characterized by a radius $R_{bin,2} \approx 0.140$ mas; it ejected only three 
VLBI components between 1994 and 1999. The two black hole sytems are separated by $\approx 1.35$ mas. 
We briefly discuss the consequences of the existence of binary black holes systems in radio quasars 
to make the link between radio quasars and GAIA.

\keywords{Astrometry - individual: 1928+738 - Galaxies: jets}}

\maketitle

\section{Introduction}
\label{sec:intro}

VLBI observations of compact radio sources show that the ejection of VLBI 
components does not follow a straight line, but wiggles. These observations 
suggest a precession of the accretion disk. By studying the observed wiggles, 
several authors raised evidences that nuclei of extragalactic radiosources contain 
binary black hole (BBH) systems (see \citet{BrRo+:01} for 0420-014, 
\citet{LoRo:05} for 3C 345, \citet{RoBr+:08} for 1803+784, \citet{RoBr+:13}  
for 1823+568 and 3C 279 and \citet{RoKa+:93}, \citet{KuGa+:14}, and 
this work for 1928+738). BBH systems can form when galaxies merge \citep{BeBl+:80} 
and the detection of BBH systems associated with nuclei of extragalactic radio sources 
can explain why extragalactic radio sources are associated with elliptical galaxies 
and why quasars (quasi stellar radio sources) represent about 5\% of the 
quasi stellar objects (QSO) \citep{BrRo+:01}. For a review concerning 
massive binary black holes systems, see \citet{CoDo:11}.

A BBH system produces three perturbations of the VLBI ejection due to
\begin{enumerate}
    \item the precession of the accretion disk,
    \item the motion of the two black holes around the center of gravity 
    of the BBH system, and
    \item the possible motion of the BBH system around 
		either a third black hole or an other BBH system. This third perturbation 
		produces a change of the VLBI jet direction. It is observed for 
		1928+738 (Fig~\ref{fig:1_1928+738_2cm03_08_Bis_3}), 3C 345 \citep{LiHo:05} and 
		3C 454.3 \citep{LiAl+:09} for instance\footnote{The BBH system can turn around 
		the center of gravity of the galaxy, however the rotation period 
		will be very large.}.
\end{enumerate}

A BBH system induces several consequences, which are that
\begin{enumerate}
    \item even if the angle between the accretion disk and the plane 
    of rotation of the BBH system is zero, the ejection does not follow 
    a straight line (due to the rotation of the black holes around the 
    center of gravity of the BBH system),
    \item the two black holes can have accretion disks with different 
    angles with the plane of rotation of the BBH system and can eject 
    VLBI components; in that case we observe two different families 
    of trajectories, i.e. trajectories ejected from two different 
		origins; a good example of a source showing two families 
		of trajectories is 3C 279 \citep{RoBr+:13}, and
    \item if the VLBI core is associated with one black hole, and if  
     the VLBI component is ejected by the second black hole, 
    there will be an offset between the VLBI core and the origin of the 
    ejection of the VLBI component; this offset will correspond to the radius 
    of the BBH system; a good example of a component ejected with a large offset 
		from the VLBI core is component C5 of 3C 279 \citep{RoBr+:13}.
\end{enumerate}

The existence of BBH systems impact several domains of astronomy and astrophysics. 
As indicated in \citet{BrRo+:01} nuclei of extragalactic radio sources which 
contain BBH systems will be good candidates to be observed with low frequency 
gravitational wave detectors like eLISA. \citet{BrRo+:01} showed that the typical 
life time of BBH systems associated with nuclei of extragalactic radio sources 
is between 5 to 10 billion years and during the final phase of the collapse there 
are observable during about 2.5 years. \citet{BrRo+:01} also estimated that the 
frequency of the collapse of BBH systems associated with extragalactic radio sources 
will be about one collapse every 2.5 years.


The existence of BBH systems in nuclei of extragalactic radio sources has also 
consequences for the realization of celestial frames. The absolute position of radio sources 
as measured by geodetic VLBI shows a standard deviation (rms) larger than 0.1 mas (see Sec. 
\ref{section:Discussion_conclusion}). This floor is partly due to the source structure, 
and constitutes a limit to the stability of a quasar-based celestial frame axes 
(e.g., \citet{FeGo+:09}; \citet{La:13}).\\

In Sec.~\ref{sec:Parameters} we will recall the parameters of the model.

In Sec.~\ref{sec:1928+738} we will give the properties of the radio source 1928+738. 

We found modeling and fitting the ejections of components C1, C5, C6, C7a, C8 and C9 
that the nucleus of 1928+738 contains at least three black holes ejecting the VLBI 
components. We shown that the three black holes belong to two BBH systems, namely 
Cg-CS and BHC6-BH4. The black hole associated with the stationary component CS ejected 
components C1 and C9, the black hole associated with the stationary component Cg ejected 
components C8 and C7a, the third black hole BHC6 which ejected C6 and C5 is not detected 
in radio. In Sec.~\ref{sec:families} we will explain to which family of 
trajectory belong the different components. 
 
To find a precise determination of the two 
BBH systems:
\begin{enumerate}
	\item we found the characteristics of the BBH system using the coordinates 
	given by \citet{KuGa+:14},
	\item we estimated the perturbation due to the slow rotation of the BBH system Cg-CS 
	around the second BBH system BHC6-BH4 assuming 
	$M_{BHC6} + M_{BH4} = (M_{CS} + M_{Cg}) /10$ and corrected the coordinates given 
	by \citep{KuGa+:14} from this perturbation, finally
	\item we used the corrected coordinates to find the final characteristics of the BBH system.
\end{enumerate}
In Sec.~\ref{sec:solution_C8_1928+738} we give the solution of the fit of component C8 after 
correction of the slow rotation of the BBH system Cg-CS 
around the second BBH system BHC6-BH4. 

In Sec.~\ref{sec:solution_C1_1928+738} we give the solution of the fit of component C1 after 
correction of the slow rotation of the BBH system Cg-CS 
around the second BBH system BHC6-BH4. 

In Sec.~\ref{sec:solution_C6_1928+738} we give the solution of the fit of component C6 after 
correction of the slow rotation of the BBH system BHC6-BH4 
around the second BBH system Cg-CS. 

Finally, in Sec.~\ref{section:Discussion_conclusion} we study the consequences of BBH systems 
in nuclei of extragalactic radio sources to link radio positions obtained from VLBI observations 
and GAIA. 

The fits of C1, C5, C6, C7a, C8 and C9 using the coordinates given by 
\citet{KuGa+:14} and the circular orbit corrections are given in the various appendices.

\section{Parameters of the model}
\label{sec:Parameters}
A VLBI component is a cloud of $e^{-}-e^{+}$ ejected relativistically. 
It corresponds to the relativistic beam in the two-fluid model. It 
follows the perturbed magnetic field lines, so its motion is not a ballistic 
motion. We will call x, y and z the coordinates of a point source component. 
For details concerning the geometry of the model, the two-fluid 
model, the perturbation of the VLBI ejection, the coordinates of the VLBI 
component see \citet{RoBr+:08} and \citet{RoBr+:13}.

The possible free
parameters of the model (for more details see \citet{RoBr+:13}). They are
\begin{itemize}
  \item $i_{o}$ the inclination angle,
    \item $\phi_{o}$ the phase of the precession at $t=0$,
    \item $\Delta\Xi$ the rotation angle in the plane perpendicular
    to the line of sight, also the asymptotic direction of the jet,
    \item $\Omega$ the opening angle of the precession cone, 
    \item $R_{o}$ the maximum amplitude of the perturbation, 
    \item $T_{p}$ the precession period of the accretion disk,
    \item $T_{d}$ the characteristic time for the damping of the beam perturbation,
    \item $M_{1}$ the mass of the black hole ejecting the radio jet,
    \item $M_{2}$ the mass of the secondary black hole,
    \item $\gamma_{c}$ the bulk Lorentz factor of the VLBI component,
    \item $\psi_{o}$ the phase of the BBH system at $t=0$,
    \item $T_{b}$ the period of the BBH system,
    \item $t_{o}$ the time of the origin of the ejection of the VLBI component,
    \item $V_{a}$ the propagation speed of the perturbations,
    \item $n_{rad}$ is the number of steps to describe the extension of the VLBI 
    component along the beam,
    \item $\Delta W$ and $\Delta N$ the possible offsets of the origin of the VLBI
    component.
\end{itemize}

As $M_{1}$ and $M_{2}$ are free parameters, the ratio $M_{1} /M_{2}$ 
is also a free parameter.

The parameter $V_a$ can be used to study the degeneracy of 
the solutions, so we can keep it constant to find the solution. The range of 
values that we study for parameter $V_a$ is $0.001 \times c \leq V_a \leq 0.45 \times c$ 
\footnote{We limit ourselves to non relativistic hydrodynamics in this model.}.

The parameter $n_{rad}$ is known when the size of the VLBI component is known.

This means that, practically, the problem we have to solve is a 15 free 
parameter problem.

We have to investigate the different possible scenarios with regard to the 
sense of the rotation of the accretion disk and the sense of the orbital 
rotation of the BBH system. These possibilities correspond to $\pm\:
\omega_{p}(t- z/V_{a})$ and $\pm \:\omega_{b}(t- z/V_{a})$. Because the
sense of the precession is always opposite to the sense of the
orbital motion \citep{Ka:97}, 
we study the two cases denoted by $+-$ and $-+$, 
where we have $\omega_{p}(t- z/V_{a})$, $-\omega_{b}(t- z/V_{a})$ and
$-\omega_{p}(t- z/V_{a})$, $\omega_{b}(t- z/V_{a})$, respectively 
($\omega_{p}$ and $\omega_{b}$ are defined by $\omega_{p} = 2 \pi /T_{p}$ and 
$\omega_{b} = 2 \pi /T_{b}$).

\section{Radio source 1928+738}
\label{sec:1928+738}
\begin{figure}[ht]
\centerline{
\includegraphics[scale=0.5, width=5.5cm,height=8cm]{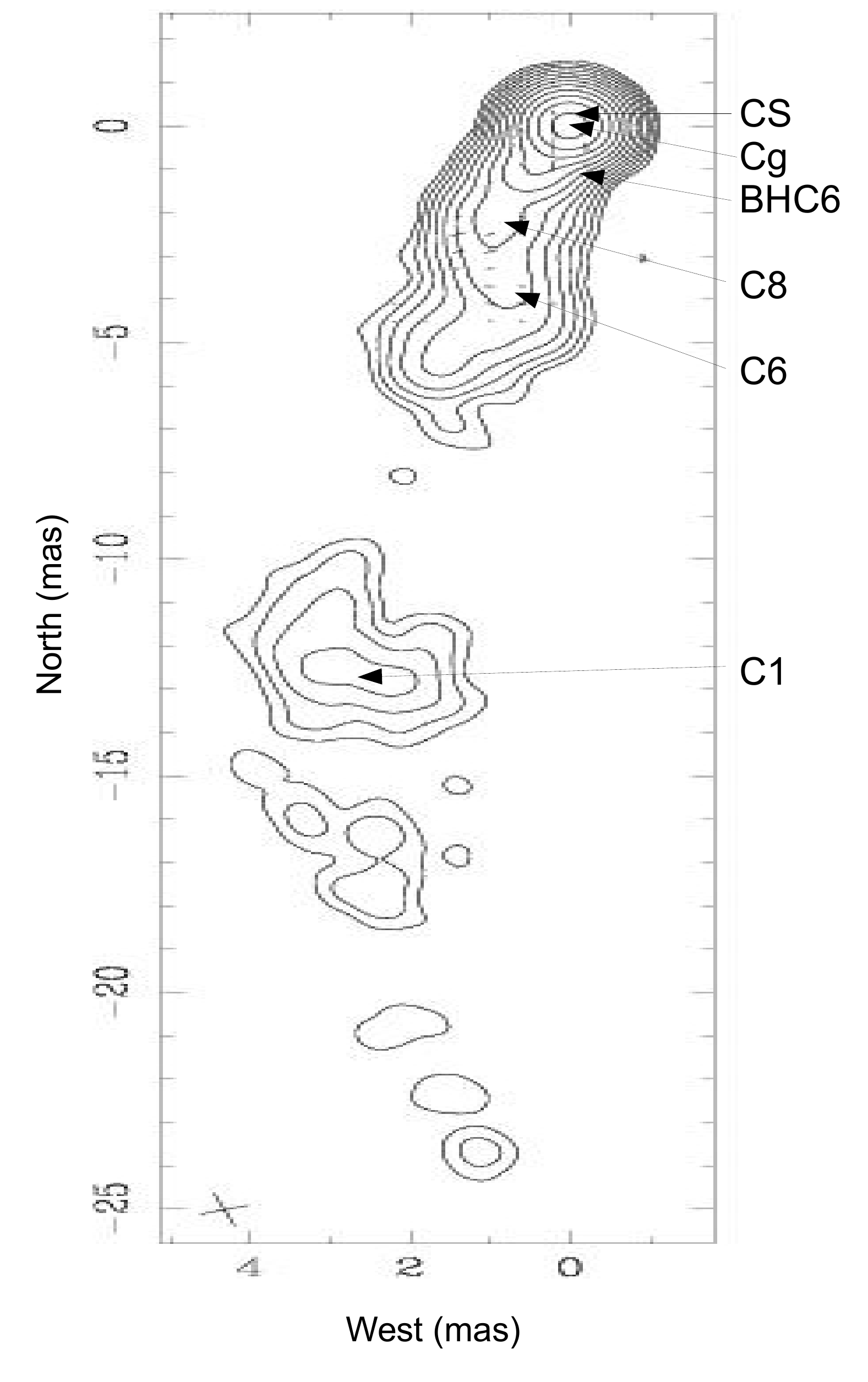}}
\caption{VLBI image of S5 1928+738 observed 28 Aug 2003 and obtained by \citet{LiHo:05}. 
The VLBI jet turns after about 10 mas and the asymptotic direction is characterized by 
$\Delta\Xi \approx 182^{\circ}$ corresponds to the direction of the large scale jet 
observed with the VLA \citep{MuBr+:93}. This long term turn is characteristic of the 
slow rotation of the BBH system around a third black hole or a second BBH system. 
We indicated the positions of the two stationary components Cg and CS which 
are associated with the two black holes of the BBH system Cg-CS, the position 
of the black hole BHC6 which ejected components C5, C6 and C7b and finally the 
positions of components C1, C6 and C8 which ejections have been fitted.}
\label{fig:1_1928+738_2cm03_08_Bis_3}
\end{figure}

The radio source S5 1928+738 is a core dominated quasar at a redshift 
of ~0.302 \citep{LaPe+:86}. It is associated with a bright optical blazar 
which magnitude is $m_{R} \approx 15$ to $16$ \citep{HeRo+:08}. 
The jet morphology is two-sided on kpc scales, the southern part being 
more pronounced \citep{MuBr+:93}. On pc scales the source is one-sided. 
The map of 1928+738 (Fig.~\ref{fig:1_1928+738_2cm03_08_Bis_3}) 
shows that the VLBI jet turns after about 10 mas \citet{LiHo:05}. 
Close to the nucleus the ejection direction is characterized by 
$\Delta \Xi \approx 163\degr$ and after about 10 mas, the VLBI jet turns 
and has an asymptotic direction characterized by $\Delta \Xi \approx 182\degr$ 
which corresponds to the large scale jet direction 
observed with the VLA \citep{MuBr+:93}. As mentioned in Sec. \ref{sec:intro}, 
the VLBI jet turn could be an indication that the nucleus of 1928+738 contains 
either three black holes or two BBH systems. We will see that components C5 and C6 
are ejected by either a third black hole or a second BBH system, so we will have 
to estimate the influence of the slow rotation of the BBH system Cg-CS around the mass
ejecting C5 and C6 and we will have to correct the coordinates given by \citet{KuGa+:14} from 
this perturbation to obtain the precise characteristics of the BBH system Cg-CS.

VLBI observations of S5 1928+738 used in the present work were taken between 1994.67 
and 2013.06 at 15 GHz in the framework of the MOJAVE Survey \citet{LiAl+:09} and \citet{LiCo+:09}. 
\citet{KuGa+:14} decomposed the brightness distribution of the jet for its components by using 
the Caltech \textsc{Difmap} software package \citep{Sh:97}. The core and the jet components 
were fitted by circular Gaussian. For details about the error calculation and component 
identification see \citet{KuGa+:14}. We will consider the VLBI core 
is associated with component Cg; \citet{KuGa+:14} consider that the VLBI core 
is associated with component CS. So the original coordinates of \citet{KuGa+:14} 
have been corrected to take into account this new origin. In this article when 
we refer to the coordinates of \citet{KuGa+:14}, they correspond to the coordinates 
corrected to take into account the origin Cg. 

\begin{figure*}[ht]
\centerline{
\includegraphics[scale=1, width=16cm,height=12cm]{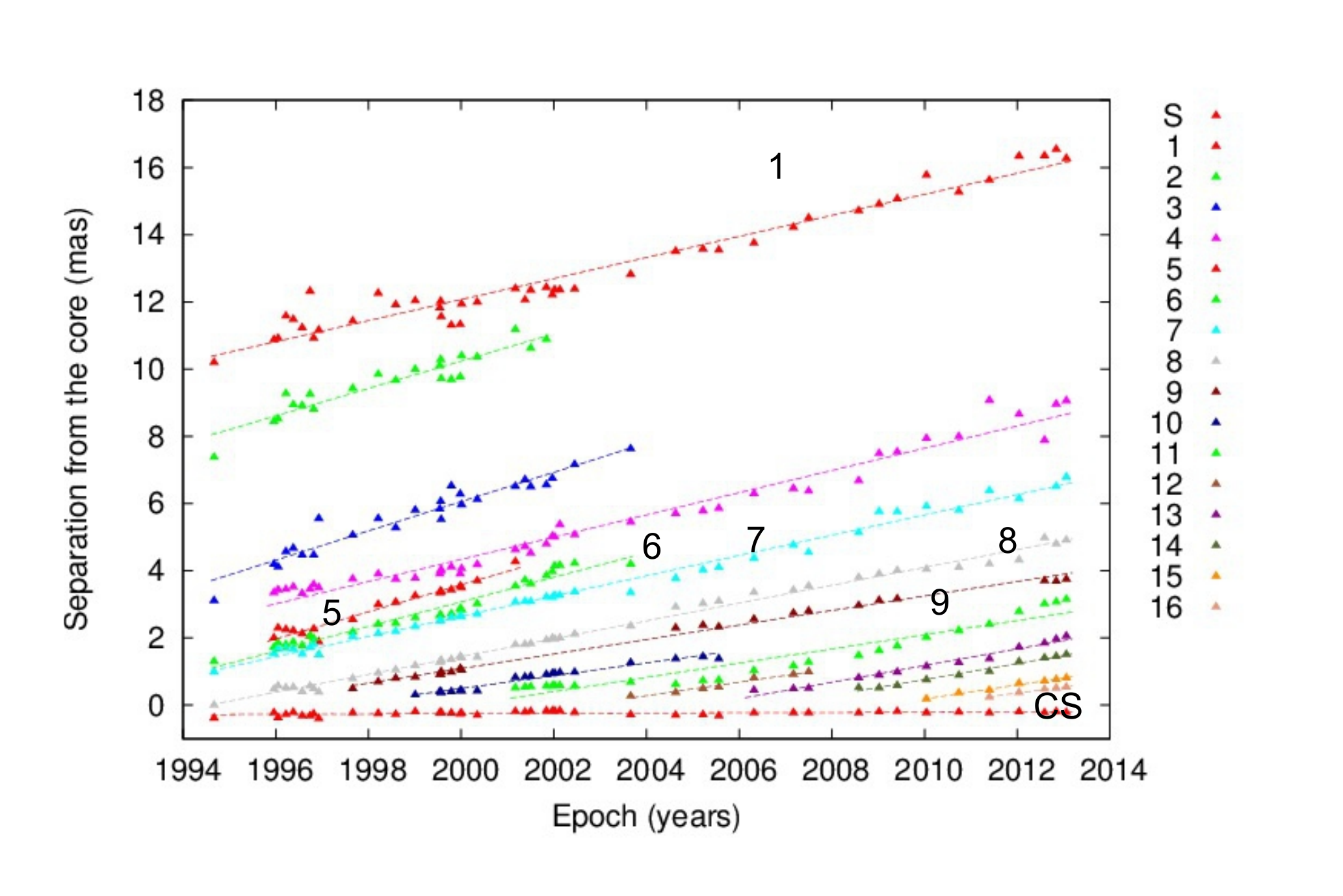}}
\caption{Separation from the core for the different VLBI 
components obtained by \citet{KuGa+:14} for the source 1928+738 using 
from MOJAVE data \citep{LiCo+:09}. We  fitted  components C1, C6 and C8 which 
have been ejected respectively by black holes Cg, CS and BHC6 and to check the 
consistency of the model found we fitted components C7a, C9 and C5 which have been 
ejected respectively by black holes Cg, CS and BHC6.}
\label{fig:3_1928+738_septime_v3_Final}
\end{figure*}

The kinematics of the jet reveals superluminal motion of its components. 
The jet components show outward motion, except the northernmost component, component CS, 
which is observed at a quasi-stationary position compared to the core. 
$16$ components appear at 15 GHz, averagely $10$-$14$ in one epoch. 
The separation from the core of the different components is shown in 
Fig~\ref{fig:3_1928+738_septime_v3_Final}.
The radio map of 1928+738, observed 28 Aug 2003, is showed in 
Fig.~\ref{fig:1_1928+738_2cm03_08_Bis_3}. 


The redshift of the source is $z_{s} \approx 0.302$, 
and using for the Hubble constant $H_{o} \approx 71$ km/s/Mpc, 
the luminosity distance of the source is $D_{l}\approx 1552$ Mpc 
and the angular distance is $D_{a} = D_{l} /(1+z)^2$. Thus 
$1$ $mas$ $ \approx 4.44$ pc.

Observations were performed at 15 GHz and the beam size
is mostly circular and equal to $Beam \approx 0.5$ $mas$. We adopted as minimum 
values of the error bars the values $(\Delta W)_{min} \approx Beam/15 \approx 34$ $\mu as$ and 
$(\Delta N)_{min} \approx Beam/15 \approx 34$ $\mu as$ for the west and north coordinates, 
i.e., when the error bars obtained from the VLBI data reduction were  
smaller than $(\Delta W)_{min}$ or $(\Delta N)_{min}$, they were enlarged to the 
minimum values (see \citet{RoBr+:13} for details concerning this choice). 

It has been suggested by \citet{LiHo:05} that the minimum values 
for the error bars should be $\approx Beam /5$, however \citet{RoBr+:13} 
shown that the correct minimum values for the error bars adopted at 15 GHz, 
are given by 
\begin{equation}
    Beam /15 \leq \Delta_{min} \leq Beam /12 \ .
\label{eq:min_error}
\end{equation}
The fit of VLBI coordinates 
of components of 3C 345 (work in progress) indicates that the adopted values 
for the minimum values of the error bars, using equation (\ref{eq:min_error}), 
are correct for frequencies between 8 GHz and 22 GHz. At lower frequencies, 
the minimum values may be higher than $Beam /12$ due to strong opacity effects 
and at 43 GHz, the minimum values are also probably higher ($\approx 20$ $\mu$as).

There are two important points concerning the minimum values used for the error bars:
\begin{enumerate}
  \item The minimum values are chosen empirically, but the adopted values are justified 
	a posteriori by comparing of the value of $\chi^{2}$ of the final solution and the 
	number of constraints used to make the fit. Indeed, the reduced $\chi^{2}$ has to be close to 1.
	\item The adopted minimum value of the error bars also includes typical errors 
	due to opacity effects, which shift the measured position at different frequencies \citep{Lo:98}.
\end{enumerate}

We will model and fit the coordinates $W(t)$ and $N(t)$ of components C8, C1 and C6 
which, as our modeling shows, have been ejected respectively by the three different black holes 
Cg, CS and BHC6 (Fig.~\ref{fig:3_1928+738_septime_v3_Final}, Sec.~\ref{sec:solution_C8_1928+738}, 
Sec.~\ref{sec:solution_C1_1928+738}, and 
Sec.~\ref{sec:solution_C6_1928+738}) and to check the consistency of the model found, 
we fit the coordinates of components C7a, C9 and C5 
which have been ejected respectively by Cg, CS and BHC6 
(see Sec.~\ref{Fit_C7a_1928+738}, Sec.~\ref{Fit_C9_1928+738} and Sec.~\ref{Fit_C5_1928+738}).

The model and the fit of components C8, C1, C6, C7a, C9 and C5 have been obtained 
with the sense of the rotation of the accretion disks $-\: \omega_{p}(t- z/V_{a})$ 
and the sense of the orbital rotation of the BBH systems $+ \:\omega_{b}(t- z/V_{a})$.

\section{Families of trajectories in the VLBI jet} 
\label{sec:families}
Modeling the ejections of components C8, C1, C6, C7a, C9 and C5, 
we found that the VLBI jet is a blend of three families of trajectories, 
i.e. the nucleus of 1928+738 contains at least three black holes ejecting VLBI components.
The first one corresponds to the VLBI components ejected by CS, 
the second one corresponds to the VLBI components ejected by Cg and 
the third one corresponds to the VLBI components ejected by the third black hole, 
we will call BHC6.  

In this section we will indicates to which family of trajectory belong the 
components. When the fit of the component has not been done we indicate the possible membership 
of the component using the trajectories, the distance from the core variations and the flux densities 
variations, and we give the time origin of the ejection. 
We use the data and component identifications from \citet{KuGa+:14}.

\begin{figure}[ht]
\centerline{
\includegraphics[scale=0.5, width=9cm,height=14cm]{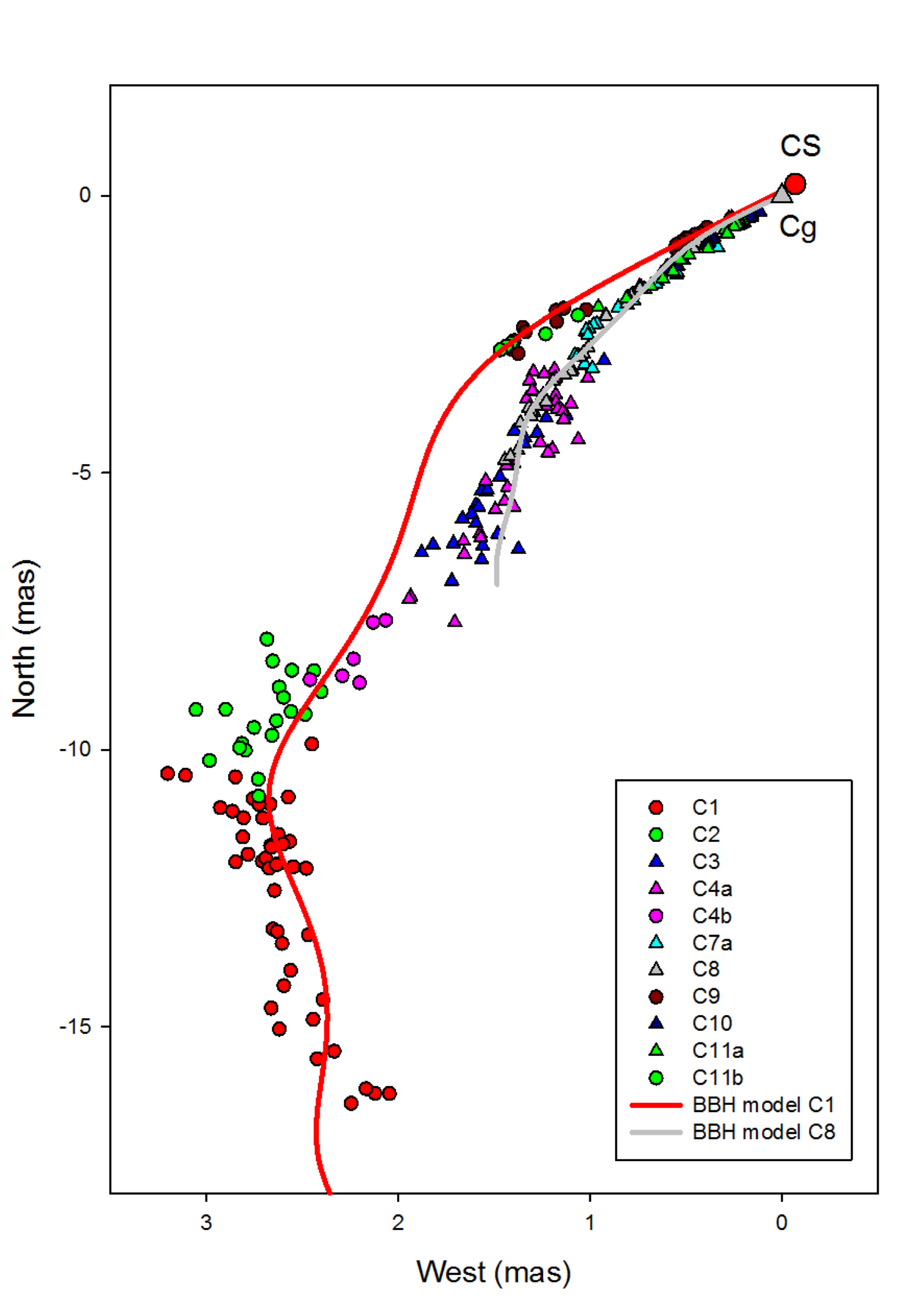}}
\caption{The two families of trajectories, associated with the VLBI 
components ejected by CS and the VLBI components ejected by Cg. 
The two families separate clearly after 1 mas. Components C1 and C9 have been 
ejected by CS and components C2, C4b and C11b have been probably ejected by CS. 
They form the first family of VLBI trajectories. Components C8 and C7a 
have been ejected by Cg and components C3, C4a, C10 and C11a have been probably ejected by Cg. 
They form the second family of VLBI trajectories. 
The VLBI coordinates are taken from \citet{KuGa+:14}.}
\label{fig:4_1928+738_Trajectories_CgCS_Bis_Final_2}
\end{figure}

Component C1: Modeling the ejection of C1, we found after the circular orbit correction that 
the mass ratio $M_{CS}/M_{Cg}$ is $M_{CS}/M_{Cg} \approx 1/3$ indicating that component C1 
cannot be ejected by the same black hole responsible for the ejection of component C8, i.e. C1 is 
not ejected by Cg but by CS (see Sec.~\ref{sec:solution_C1_1928+738}, 
Fig~\ref{fig:13_SOLUTION_C1_Distance+Vap_Time_Off}, Fig~\ref{fig:14_SOLUTION_C1_Xt+Yt_Time_Off} 
and Fig.~\ref{fig:4_1928+738_Trajectories_CgCS_Bis_Final_2}). 
So, C1 define the first family of trajectories and has been ejected 
by the black hole associated with CS which coordinates 
are $X_{CS} \approx -0.07$ mas and $Y_{CS} \approx +0.21$ mas and $t_{o} \approx 1967$.

Component C2: follows the same trajectory than C1, so has probably been ejected by CS, 
and $t_{o} \approx 1975$ (Fig.~\ref{fig:4_1928+738_Trajectories_CgCS_Bis_Final_2}).

Component C3: follows the same trajectory than C8, so has probably been ejected by 
the black hole associated with Cg which coordinates are $X_{CS} = 0.0$ mas and $Y_{CS}  +0.0$ mas  
and $t_{o} \approx 1986$ (Fig.~\ref{fig:4_1928+738_Trajectories_CgCS_Bis_Final_2}).

Component C4: is possibly a blend of two components C4a and C4b. 
The beginning of the trajectory, C4a i.e. from 1996 to 2008.6, is the same than C8, 
so it has probably been ejected by Cg and $t_{o} \approx 1987$. The end of the trajectory, 
C4b i.e. the last eight points from 2008.9 to 2013.1, is the same than C1, and 
$t_{o} \approx 1990$ (Fig.~\ref{fig:4_1928+738_Trajectories_CgCS_Bis_Final_2}).

\begin{figure}[ht]
\centerline{
\includegraphics[scale=0.5, width=8cm,height=6cm]{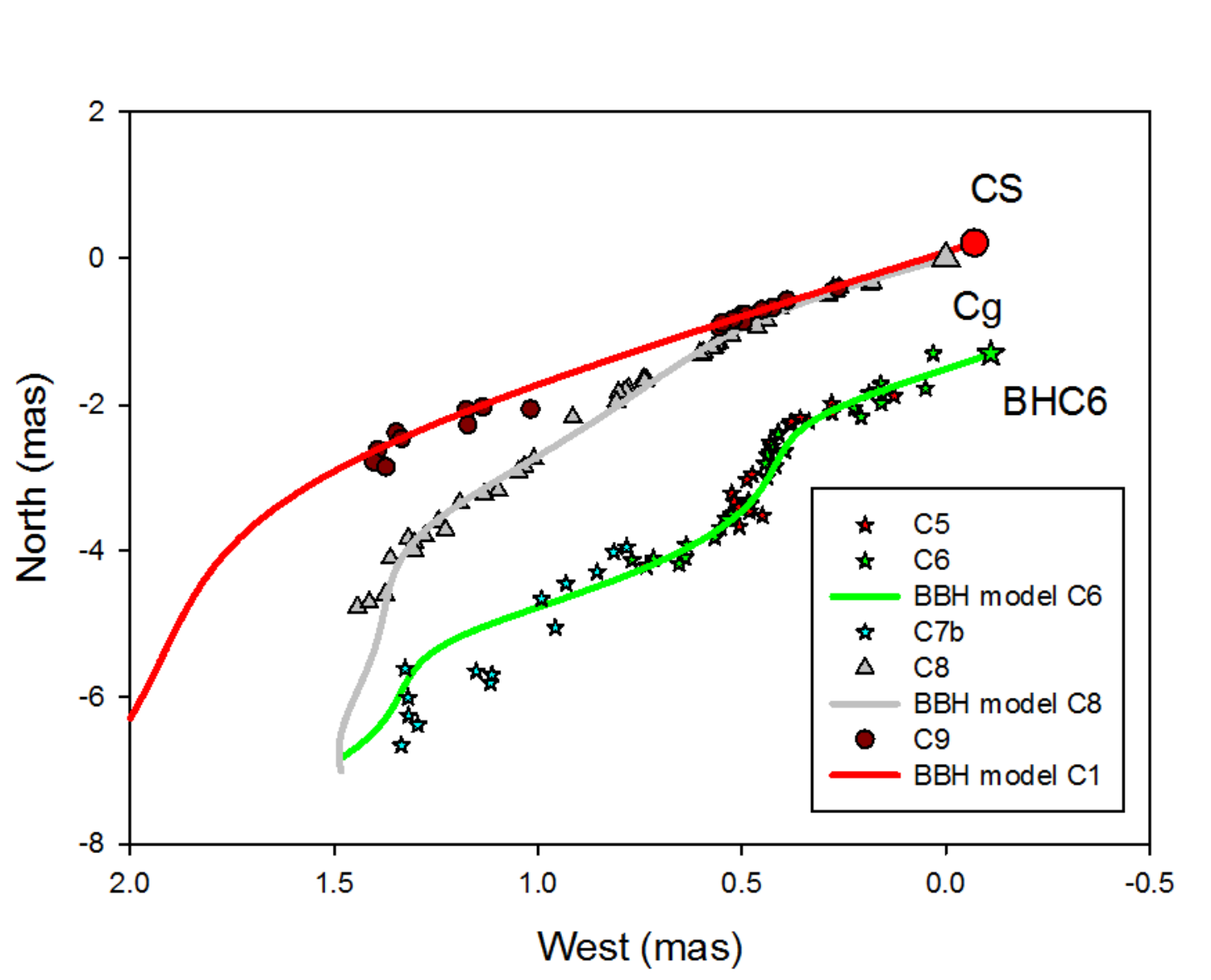}}
\caption{The family of trajectories, associated with the VLBI 
components ejected by the third black hole BHC6. Components C5 and C6, have been ejected 
by BHC6 and C7b has been probably ejected by BHC6. They form the third family of VLBI trajectories. 
The VLBI coordinates are taken from \citet{KuGa+:14}.}
\label{fig:5_1928+738_Trajectories_BH3_Bis_Final_2}
\end{figure}

Component C5: follows the same trajectory than C6 which is different from C1 
and C8.
To check that component C5 belongs to the family 
of components ejected by the black hole BHC6 and the consistency 
of the model found, we used the characteristics 
of the BBH system BHC6-BH4 and the characteristics 
of the geometrical parameters of the trajectory 
of C6, to fit the coordinates of components C5  
(see Sec.~\ref{Fit_C5_1928+738}, Fig.~\ref{fig:19_comp5_v4_2_Traj_New} and 
Fig.~\ref{fig:5_1928+738_Trajectories_BH3_Bis_Final_2}). 
So, C5 is ejected by the black hole BHC6, so it belongs to the third family of trajectory,  
and $t_{o} \approx 1991$.

\begin{figure}[ht]
\centerline{
\includegraphics[scale=0.5, width=8cm,height=6cm]{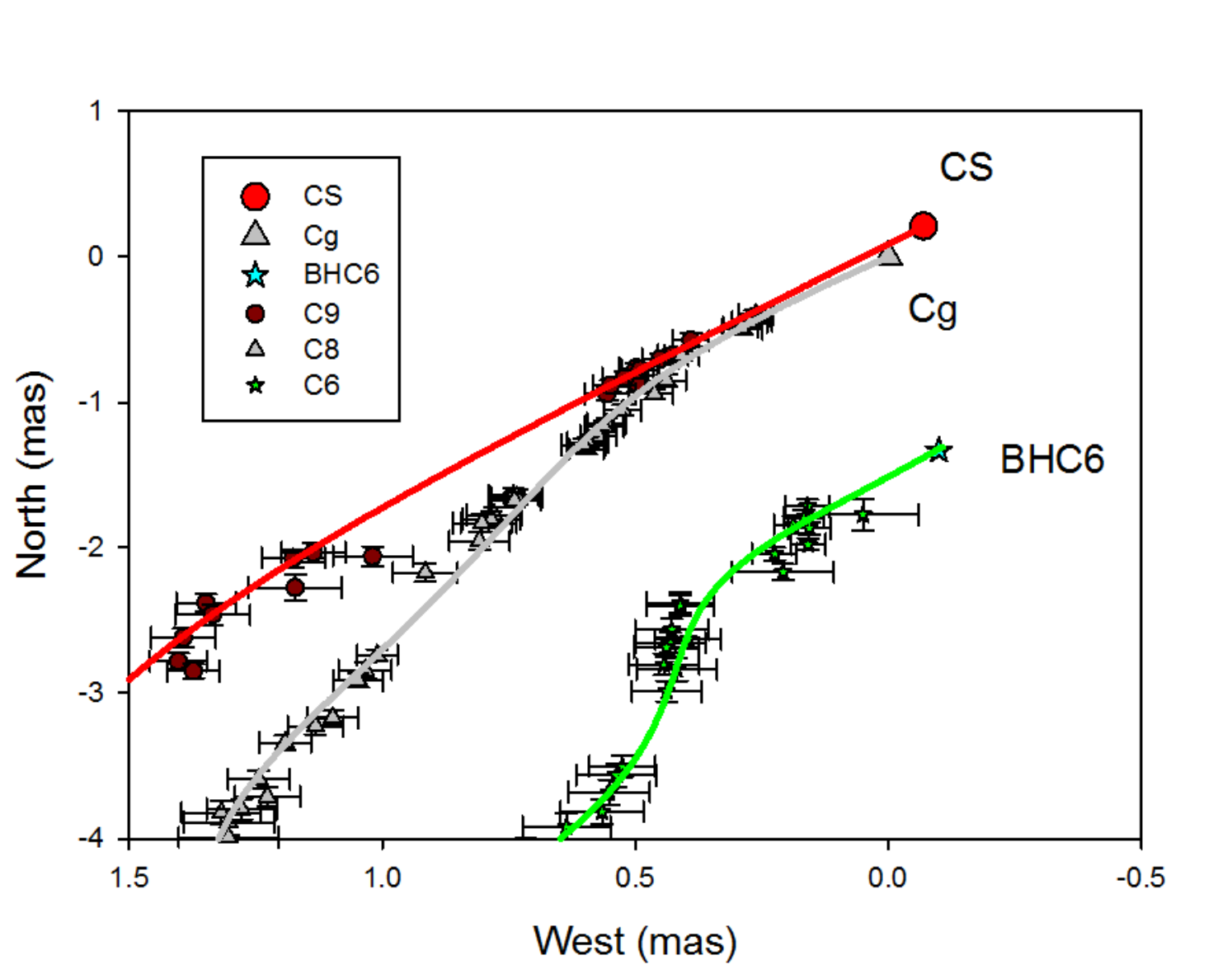}}
\caption{Structure of the nucleus of 1928+738. The nucleus of 1928+738 contains 
two BBH systems separated by $\approx 1.35$ mas $\approx 6$ pc. The first BBH 
system is constituted by components Cg and CS and has a size $R_{bin,1} \approx 220$ 
$\mu$as $\approx 0.98$ pc. The second BBH system is constituted by BHC6 and BH4 
which are not detected in radio and has a size $R_{bin,2} \approx 140$ $\mu$as 
$\approx 0.62$ pc. The position of BH4 is unknown.}
\label{fig:7_1928+738_Nucleus_Structure_Bis_Final}
\end{figure}

Component C6 : define the third family of trajectories, i.e. is ejected by 
the third black hole, BHC6, which coordinates are $X_{BHC6} \approx -0.10$ mas 
and $Y_{BHC6} \approx -1.30$ mas (see Sec.~\ref{sec:solution_C6_1928+738}, 
Fig.~\ref{fig:16_Sol2_Comp6_Distance+Vap_Time_New1_BH3_Off}, Fig.~\ref{fig:17_Sol2_Comp6_Xt+Yt_New1_BH3_Off}
Fig.~\ref{fig:5_1928+738_Trajectories_BH3_Bis_Final_2} and Fig.~\ref{fig:7_1928+738_Nucleus_Structure_Bis_Final}), 
and $t_{o} \approx 1994.5$.

Component C7 : is a blend of two components C7a and C7b. The beginning of the trajectory 
(from 1994.5 to 2002.5), C7a, is the same than C8. 
To check that component C7a belongs to the family 
of components ejected by the black hole Cg and the consistency 
of the model found, we used the characteristics 
of the BBH system Cg-CS and the characteristics 
of the geometrical parameters of the trajectory 
of C8, to fit the coordinates of components C7a 
(see Sec.~\ref{Fit_C7a_1928+738}, Fig.~\ref{fig:12a_comp7a_v4_Traj_New1} 
and Fig.~\ref{fig:4_1928+738_Trajectories_CgCS_Bis_Final_2}).  
So, C7a belongs to the second family 
of trajectories, i.e. has been ejected by Cg 
and $t_{o} \approx 1992$ . 
The end of the trajectory, C7b, is the same that C5 and C6, so C7b 
has been probably ejected by BHC6, and $t_{o} \approx 1998.5$ 
(Fig.~\ref{fig:5_1928+738_Trajectories_BH3_Bis_Final_2}).

Component C8: Modeling the ejection of C8, we found after the circular orbit correction that 
the mass ratio $M_{Cg}/M_{CS}$ is $M_{Cg}/M_{CS} \approx 3$  and there no offset of the origin of the 
ejection, indicating that component C8 cannot be ejected by the same black hole responsible 
for the ejection of component C1, i.e. C8 is 
not ejected by CS but by Cg  (see Sec.~\ref{sec:solution_C8_1928+738}, 
Fig.~\ref{fig:10_Comp8_Distance+Vap_Time_15G_New4}, 
Fig.~\ref{fig:11_Comp8_Xt+Yt_15G_New4} and Fig.~\ref{fig:4_1928+738_Trajectories_CgCS_Bis_Final_2}).  
So, C8 define the second family of trajectories and 
has been ejected by Cg and $t_{o} \approx 1993.8$ .

Component C9: follows the same trajectory than C1. 
To check that component C9 belongs to the family 
of components ejected by the black hole CS and the consistency 
of the model found, we used the characteristics 
of the BBH system Cg-CS and the characteristics 
of the geometrical parameters of the trajectory 
of C1, to fit the coordinates of components C9 
(see Sec.~\ref{Fit_C9_1928+738}, 
Fig.~\ref{fig:15a_comp9_v4_Traj_2_New_Off_CS_Short} and Fig.~\ref{fig:4_1928+738_Trajectories_CgCS_Bis_Final_2}). 
So, C9 belongs to the first family of trajectories, i.e. has been ejected by CS, 
and $t_{o} \approx 1995$.

Component C10: follows the same trajectory than C8, has been probably ejected by Cg, 
and $t_{o} \approx 1997.8$ (Fig.~\ref{fig:4_1928+738_Trajectories_CgCS_Bis_Final_2}).

Component C11: is a blend of two components, C11a and C11b. The beginning of the trajectory, C11a, 
follows the same trajectory than C8, has been probably ejected by Cg, 
and $t_{o} \approx 1998$ (Fig.~\ref{fig:4_1928+738_Trajectories_CgCS_Bis_Final_2}). 
The end of the trajectory, C11b, follows the same trajectory than C9, has been probably ejected by CS, 
and $t_{o} \approx 2003$ (Fig.~\ref{fig:4_1928+738_Trajectories_CgCS_Bis_Final_2}). 

For the next components, their trajectories are not long enough to determine if their trajectories 
are similar to C1 or C8. However, none of then are ejected by BHC6.

Component C12: has been ejected by Cg or CS, and $t_{o} \approx 2002.5$.

Component C13: may be a blend of two components that have been ejected by Cg or CS, and 
$2005 \leq t_{o} \leq 2007$.

Component C14: has been ejected by Cg or CS, and $t_{o} \approx 2006.5$.

Component C15: has been ejected by Cg or CS, and $t_{o} \approx 2009$.

Component C16: has been ejected by Cg or CS, and $t_{o} \approx 2010$.\\

We found that the VLBI components follow three different families of trajectories, 
i.e. the nucleus of 1928+738 contains at least three black holes. 
However, as we will see, the fit of the ejection of C6 cannot be explained 
simply a third black hole but indicates that C6 is ejected by a second BBH system. 
So the nucleus of 1928+738 contains 2 BBH systems.
Let us call
\begin{enumerate}
	\item Cg and CS the two black holes of the first BBH system and
	\item BHC6 and BH4 the two black holes of the second BBH system.
\end{enumerate}

As indicated in Sec.~\ref{sec:intro} the BBH system Cg-CS turns around 
the second BBH system BHC6-BH4, so we will have to correct this slow rotation 
to make a precise determination of the two BBH systems parameters.

We will model and fit the coordinates of components C8, C1 and C6 and we will be 
able to deduce the characteristics of the two BBH systems (see 
Sec.~\ref{sec:solution_C8_1928+738}, Sec.~\ref{sec:solution_C1_1928+738}, 
Sec.~\ref{sec:solution_C6_1928+738} and Sec.~\ref{section:Discussion_conclusion}). 
To check the consistency of the model found, we will use the characteristics 
of the two BBH systems and the geometrical parameters of the trajectories of 
C8, C1 and C6 to fit the coordinates of components C7a, C9 and C5 
which have been ejected respectively by Cg, CS and BHC6 
(see Sec.~\ref{Fit_C7a_1928+738}, Sec.~\ref{Fit_C9_1928+738} and Sec~\ref{Fit_C5_1928+738}).
\\

\begin{figure}[ht]
\centerline{
\includegraphics[scale=0.5, width=8cm,height=6cm]{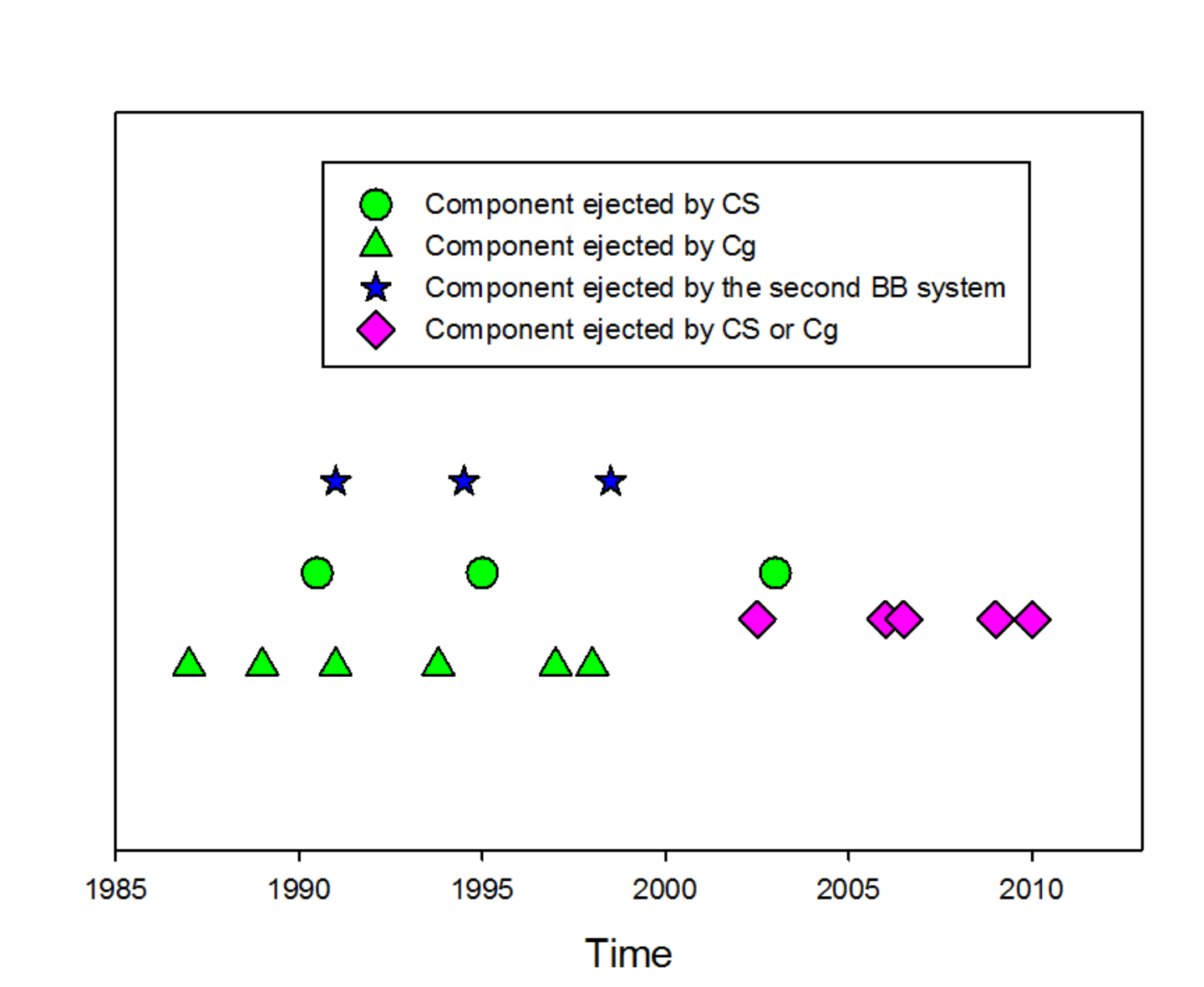}}
\caption{Time origin of the component ejected by the three black holes of the nucleus of 
1928+738 during the period from 1985 to 2015.}
\label{fig:6_1928+738_Birth_of_Components_2_Bis}
\end{figure}

We plotted in Fig.~\ref{fig:6_1928+738_Birth_of_Components_2_Bis} the time origin 
of the component ejected by three black holes of the nucleus of 1928+738 during 
the period from 1985 to 2015. We find that
\begin{itemize}
	\item there is no obvious periodicity in the ejection time of the VLBI components,
	\item black hole BHC6, ejected components only during the period from 1991 to 1999,
	\item black hole Cg ejects about two time more components 
	than black hole CS.
\end{itemize}

\section{Fit of component C8}
\label{sec:solution_C8_1928+738}

To obtain a precise determination of the characteristics of the BBH system 
ejecting component C8, 
\begin{enumerate}
	\item we found the characteristics of the BBH system Cg-CS using the coordinates 
	of C8 given by \citet{KuGa+:14}, see Sec.~\ref{Fit_C8_1928+738},
	\item we estimated the perturbation due to the slow rotation of the BBH system Cg-CS 
	around the second BBH system BHC6-BH4 assuming 
	$M_{BHC6} + M_{BH4} = (M_{CS} + M_{Cg}) /10$ and corrected the coordinates given 
	by \citep{KuGa+:14} from this perturbation, see details in Sec.~\ref{COC_C8_1928+738}, finally
	\item we used the corrected coordinates to find the final characteristics of the BBH system Cg-CS.
\end{enumerate}

Here we present the solution to the fit of the coordinates of C8 which have been 
corrected from the slow motion of the BBH system Cg-CS around the BBH system BHC6-BH4.\\

The main characteristics of the BBH system ejecting C8 are that
\begin{itemize}
    \item the VLBI component C8 is ejected by the VLBI core, i.e. component Cg 
		(there in no indication of an offset of the origin of the ejection),
    \item the two black holes are associated with the components CS and Cg,
    \item the radius of the BBH system is $R_{bin} \approx 220$ $\mu as$ $\approx 0.98$ $pc$,
    \item the ratio $M_{Cg}/M_{CS}$ is $\approx 3$, which is the inverse of the mass ratio found 
		fitting the coordinates of C1 (Sec.~\ref{sec:solution_C1_1928+738}) and
		\item the ratio $T_{p}/T_{b}$ is $\approx 107$ .
\end{itemize}

The ratio $M_{Cg}/M_{CS}$ is a free parameter of the model and is determinated 
by the fit of the coordinates of C8.\\

We find also that
\begin{itemize}
    \item the inclination angle is $i_{o} \approx 18.5^{\circ}$,
		\item the asymptotic ejection direction is $\Xi \approx 165^{\circ}$, 
    \item the angle between the accretion disk and the rotation plane 
    of the BBH system is $\Omega \approx 2.7^{\circ}$,
    \item the bulk Lorentz factor of the VLBI component is $\gamma_{c} \approx 5.9$, and
    \item the origin of the ejection of the VLBI component is $t_{o} \approx 1993.8$.
\end{itemize}

Compared to the first solution found in Sec.~\ref{Fit_C8_1928+738}, 
this new solution is characterized by a smaller inclination angle, a smaller 
mass ratio $M_{Cg}/M_{CS}$, a smaller angle between the accretion disk and 
the rotation plane of the BBH system and an asymptotic direction of the VLBI jet, 
$\Delta \Xi \approx 165\degr$ indicating that the long term turn of the VLBI jet 
observed at about 10 mas is due to the rotation of  the BBH system Cg-CS around 
the BBH system BHC6-BH4.

\subsection{Determining the family of solutions}
For the inclination angle previously found, i.e., 
$i_{o}\approx 18.5^{\circ}$, $T_{p}/T_{b} \approx 107$, $M_{Cg}/M_{CS} \approx 3$, 
and $R_{bin} \approx 220$ $\mu as$, we gradually varied $V_{a}$ between $0.001$ c 
and $0.45$ c. The function $\chi^{2}(V_{a})$ remained constant,  
indicating a degeneracy of the solution. We deduced the 
range of variation of the BBH system parameters. They are given in Table 1.

\begin{center}
Table 1 : Ranges for the BBH system parameters ejecting C8\medskip%

\begin{tabular}
[c]{c||c|c}\hline
$V_{a}$                    & $0.001 \: c$                         & $0.45 \: c$                            \\\hline
$T_{p}(V_{a})$             & $\approx 17800000$ yr                & $\approx 21400$ yr                     \\\hline
$T_{b}(V_{a})$             & $\approx 166000$ yr                  & $\approx 199$ yr                       \\\hline
$(M_{Cg}+ M_{CS})(V_{a})$  & $\approx 3 \; 10^{5}$ $M_{\odot}$    & $\approx 2.0 \; 10^{11}$ $M_{\odot}$   \\\hline
\end{tabular}
\end{center}

Table 1 provides the range of the BBH system parameters ejecting C8. To obtain the final range of the 
two BBH systems Cg-CS and BHC6-BH4 one has to make the intersection of the ranges of BBH sytems parameters 
found after the fits of C8, C1 and C6 (this is done in Sec.~\ref{section:Discussion_conclusion}).

\subsection{Determining the size of the accretion disk}

From the knowledge of the mass ratio $M_{Cg}/M_{CS} \approx 3$ and 
the ratio $T_{p}/T_{b} \approx 107$, we calculated in the previous section 
the mass of the ejecting black hole $M_{Cg}$, the orbital period $T_{b}$, 
and the precession period $T_{p}$ for each value of $V_{a}$.

The rotation period of the accretion disk, $T_{disk}$, is given by 
\citep{BrRo+:01}

\begin{equation}
    T_{disk} \approx \frac{4}{3}\frac{M_{Cg}+M_{CS}}{M_{CS}} T_{b} \frac{T_{b}}{T_{p}} \ .
    \label{eq:Tdisk}
\end{equation}

Thus we calculated the rotation period of the accretion disk, and 
assuming that the mass of the accretion disk is $M_{disk} \ll M_{Cg}$, 
the size of the accretion disk $R_{disk}$ is 
\begin{equation}
    R_{disk} \approx \left(\frac{T_{disk}^{2}}{4\pi^{2}}GM_{Cg}\right)^{1/3} \ .
    \label{eq:Rdisk}
\end{equation}

We found that the size of the accretion disk does not depend on $V_{a}$ and is 
$R_{disk}  \approx 0.027 \; mas \approx 0.120 \; pc $.

\section{Fit of component C1}
\label{sec:solution_C1_1928+738}

To obtain a precise determination of the characteristics of the BBH system 
ejecting component C1, 
\begin{enumerate}
	\item we found the characteristics of the BBH system Cg-CS using the coordinates 
	of C1 given by \citet{KuGa+:14}, see Sec.~\ref{Fit_C1_1928+738},
	\item we estimated the perturbation due to the slow rotation of the BBH system Cg-CS 
	around the second BBH system BHC6-BH4 assuming 
	$M_{BHC6} + M_{BH4} = (M_{CS} + M_{Cg}) /10$ and corrected the coordinates given 
	by \citep{KuGa+:14} from this perturbation, see see details in Sec.~\ref{COC_C1_1928+738}, finally
	\item we used the corrected coordinates to find the final characteristics of the BBH system Cg-CS.
\end{enumerate}

Here we present the solution to the fit of the coordinates of C1 which have been 
corrected from the slow motion of the BBH system Cg-CS around the BBH system BHC6-BH4.\\

The main characteristics of the BBH system ejecting C1 are that
\begin{itemize}
    \item the VLBI component C1 is not ejected by the VLBI core Cg, but by component CS 
		(there is a weak indication of an offset of the origin of the VLBI ejection 
		in the direction of CS, this weak indication is due to the lack of observations of C1 
		for the beginning of the trajectory),
		\item the coordinates of CS are $X_{CS} \approx -0.07$ mas and $Y_{CS} \approx +0.21$ mas,
    \item the radius of the BBH system is $R_{bin} \approx 220$ $\mu as$ $\approx 0.98$ $pc$,
    \item the ratio $M_{CS}/M_{Cg}$ is $\approx 1/3$, which is the inverse of the mass ratio found 
		fitting the coordinates of C8 (Sec.~\ref{sec:solution_C8_1928+738}) and
		\item the ratio $T_{p}/T_{b}$ is $\approx 31$ .
\end{itemize}

The ratio $M_{CS}/M_{Cg}$ is a free parameter of the model and the value $M_{CS}/M_{Cg} \approx 1/3$ 
comes from the fit of the coordinates of C1. 
The fact that the fit of C1 provides a mass ratio $M_{CS}/M_{Cg} \approx 1/3$ which is the inverse 
of the mass ratio $M_{Cg}/M_{CS} \approx 3$ obtained from the fit of component C8 shows that components 
C1 and C8 are not ejected by the same black hole. This result shows the consistency of the method and 
it is remarkable to find this result with only parts of the complete trajectories of C1 and C8.\\

We find that
\begin{itemize}
    \item the inclination angle is $i_{o} \approx 19^{\circ}$,
		\item the asymptotic ejection direction is $\Xi \approx 162^{\circ}$, 
    \item the angle between the accretion disk and the rotation plane 
    of the BBH system is $\Omega \approx 2.4^{\circ}$,
    \item the bulk Lorentz factor of the VLBI component is $\gamma_{c} \approx 10.2$, and
    \item the origin of the ejection of the VLBI component is $t_{o} \approx 1966.2$.
\end{itemize}

Compared to the first solution found in Sec.~\ref{Fit_C1_1928+738}, 
this new solution is characterized by a smaller inclination angle, a smaller angle 
between the accretion disk and the rotation plane of the BBH system, 
and an asymptotic direction of the VLBI jet, $\Delta \Xi \approx 162\degr$ 
indicating that the long term turn of the VLBI jet observed at about 10 mas 
is due to the rotation of the BBH system Cg-CS around the BBH system BHC6-BH4.

\subsection{Determining the family of solutions}
For the inclination angle previously found, i.e., 
$i_{o}\approx 19^{\circ}$, $T_{p}/T_{b} \approx 31$, $M_{CS}/M_{Cg} \approx 1/3$, 
and $R_{bin} \approx 220$ $\mu as$, we gradually varied $V_{a}$ between $0.001$ c 
and $0.45$ c. The function $\chi^{2}(V_{a})$ remained constant,  
indicating a degeneracy of the solution. We deduced the 
range of variation of the BBH system parameters. They are given in Table 2.

\begin{center}
Table 2 : Ranges for the BBH system parameters ejecting C1\medskip%

\begin{tabular}
[c]{c||c|c}\hline
$V_{a}$                    & $0.001 \: c$                           & $0.45 \: c$                            \\\hline
$T_{p}(V_{a})$             & $\approx 6700000$ yr                   & $\approx 8100$ yr                     \\\hline
$T_{b}(V_{a})$             & $\approx 220000$ yr                    & $\approx 265$ yr                       \\\hline
$(M_{Cg}+ M_{CS})(V_{a})$  & $\approx 1.7 \; 10^{5}$ $M_{\odot}$    & $\approx 1.2 \; 10^{11}$ $M_{\odot}$   \\\hline
\end{tabular}
\end{center}

Table 2 provides the range of the BBH system parameters ejecting C1. To obtain the final range of the 
two BBH systems Cg-CS and BHC6-BH4 one has to make the intersection of the ranges of BBH systems parameters 
found after the fits of C8, C1 and C6 (this is done in Sec.~\ref{section:Discussion_conclusion}).

\subsection{Determining the size of the accretion disk}

From the knowledge of the mass ratio $M_{Cs}/M_{Cg} \approx 1/3$ and 
the ratio $T_{p}/T_{b} \approx 31$, we calculated in the previous section 
the mass of the ejecting black hole $M_{CS}$, the orbital period $T_{b}$, 
and the precession period $T_{p}$ for each value of $V_{a}$.

The rotation period of the accretion disk, $T_{disk}$, is given by 
(\ref{eq:Tdisk}). Thus we calculated the rotation period of the accretion disk, and 
assuming that the mass of the accretion disk is $M_{disk} \ll M_{CS}$, 
the size of the accretion disk is given by (\ref{eq:Rdisk}). 
We found that the size of the accretion disk, does not depend on $V_{a}$ and is 
$R_{disk}  \approx 0.021 \; mas \approx 0.093 \; pc $.

\section{Fit of component C6}
\label{sec:solution_C6_1928+738}

The component C6 is not ejected by CS or Cg but is ejected by a third 
black hole which belongs to a second BBH system. Let us call BHC6 the black hole 
ejecting component C6 and BH4 the fourth black hole. 
If we assume that C6 is ejected by a single black hole, we applied the 
precession model and we studied the solution $\chi^{2}(i_{o})$ in the interval 
$2^{\circ} \leq i_{o} \leq 50^{\circ}$, we found that 
\begin{enumerate}
  \item there exist solutions with $\gamma < 30$ only in the interval 
	$2^{\circ} \leq i_{o} \leq 17^{\circ}$ (see Fig.~\ref{fig:16a_Chi2+Gamma_io_C6_Precession}) and
	\item the solution with $\gamma < 30$ is a mirage solution, 
	i.e. the curve $\chi^{2}(i_{o})$ is convex 	and it does not show a minimum; 
	moreover the bulk Lorentz factor $\gamma$ diverges when 
	$i_{o} \rightarrow 17^{\circ}$ (see Fig.~\ref{fig:16_Chi2+Gamma_io_C6_Precession}).
\end{enumerate}
see more details in Sec.~\ref{Fit_C6_1928+738}.

To obtain a precise determination of the characteristics of the BBH system 
ejecting component C6, 
\begin{enumerate}
	\item we found the characteristics of the BBH system BHC6-BH4 using the coordinates 
	of C6 given by \citep{KuGa+:14}, see Sec.~\ref{Fit_C6_1928+738},
	\item we estimated the perturbation due to the slow rotation of the BBH system BHC6-BH4 around the second 
	BBH system Cg-CS assuming $M_{BHC6} + M_{BH4} = (M_{CS} + M_{Cg}) /10$ and 
	corrected the coordinates given by \citep{KuGa+:14} from this perturbation, 
	see see details in Sec.~\ref{COC_C6_1928+738}, finally
	\item we used the corrected coordinates to find the final characteristics of the BBH system BHC6-BH4.
\end{enumerate}

The main characteristics of the solution of the BBH system 
ejecting C6 are that
\begin{itemize}
    \item the coordinates of BHC6 are $X_{BHC6} \approx -0.11$ mas and $Y_{BHC6} \approx -1.30$ mas 
		(assuming that the origin is associated with Cg),
    \item none of the two black holes are associated with a stationary VLBI component, 
		i.e. they are not strong sources,
    \item the radius of the BBH system is $R_{bin} \approx 140$ $\mu as$ $\approx 0.62$ $pc$,
    \item calling $M_{BHC6}$ the mass of the black hole ejecting C6 and $M_{BH4}$ the mass of 
		the other black hole, the ratio $M_{BHC6}/M_{BH4}$ is $\approx 0.12$ , and
		\item the ratio $T_{p}/T_{b}$ is $\approx 50$ .
\end{itemize}

We find that
\begin{itemize}
    \item the inclination angle is $i_{o} \approx 23^{\circ}$,
		\item the asymptotic ejection direction is $\Xi \approx 165^{\circ}$, 
    \item the angle between the accretion disk and the rotation plane 
    of the BBH system is $\Omega \approx 1.9^{\circ}$,
    \item the bulk Lorentz factor of the VLBI component is $\gamma_{c} \approx 5.7$, and
    \item the origin of the ejection of the VLBI component is $t_{o} \approx 1994.5$.
\end{itemize}

Compared to the first solution found in Sec.~\ref{Fit_C6_1928+738}, 
this new solution is characterized by a similar inclination angle, a smaller angle 
between the accretion disk and the rotation plane of the BBH system, 
and a smaller mass ratio $M_{BHC6}/M_{BH4}$.

\subsection{Determining the family of solutions}
For the inclination angle previously found, i.e., 
$i_{o}\approx 23^{\circ}$, $T_{p}/T_{b} \approx 50$, $M_{BHC6}/M_{BH4} \approx 0.12$, 
and $R_{bin} \approx 140$ $\mu as$, we gradually varied $V_{a}$ between $0.001$ c 
and $0.45$ c. The function $\chi^{2}(V_{a})$ remained constant,  
indicating a degeneracy of the solution. We deduced the 
range of variation of the BBH system parameters. They are given in Table 3.

\begin{center}
Table 3 : Ranges for the BBH system parameters ejecting C6\medskip%

\begin{tabular}
[c]{c||c|c}\hline
$V_{a}$                       & $0.001 \: c$                           & $0.45 \: c$                            \\\hline
$T_{p}(V_{a})$                & $\approx 4300000$ yr                   & $\approx 5200$ yr                      \\\hline
$T_{b}(V_{a})$                & $\approx 84600$ yr                     & $\approx 103$ yr                        \\\hline
$(M_{BHC6}+ M_{BH4})(V_{a})$  & $\approx 2.9 \; 10^{5}$ $M_{\odot}$    & $\approx 2.0 \; 10^{11}$ $M_{\odot}$   \\\hline
\end{tabular}
\end{center}

Table 3 provides the range of the BBH system parameters ejecting C6. To obtain the final range of the 
two BBH systems Cg-CS and BHC6-BH4 one has to make the intersection of the ranges of BBH systems parameters 
found after the fits of C8, C1 and C6 (this is done in Sec.~\ref{section:Discussion_conclusion}).

\subsection{Determining the size of the accretion disk}

From the knowledge of the mass ratio $M_{BHC6}/M_{BH4} \approx 0.12$ and 
the ratio $T_{p}/T_{b} \approx 50$, we calculated in the previous section 
the mass of the ejecting black hole $M_{BHC6}$, the orbital period $T_{b}$, 
and the precession period $T_{p}$ for each value of $V_{a}$.

The rotation period of the accretion disk, $T_{disk}$, is given by 
(\ref{eq:Tdisk}). Thus we calculated the rotation period of the accretion disk, and 
assuming that the mass of the accretion disk is $M_{disk} \ll M_{BHC6}$, 
the size of the accretion disk is given by (\ref{eq:Rdisk}). 
We found that the size of the accretion disk, does not depend on $V_{a}$ and is 
$R_{disk}  \approx 0.006 \; mas \approx 0.027 \; pc $.

\section{Discussion and conclusion} 
\label{section:Discussion_conclusion}
Modeling the ejections of components C8, C1, C6, C7a, C9 and C5, 
we found that the VLBI components follow three different families of trajectories, 
i.e. the nucleus of 1928+738 contains at least three black holes. The fit of the 
ejection of C6 cannot be explained simply by a third black hole but indicates that C6 
is ejected by a second BBH system, so the nucleus of 1928+738 contains 2 BBH systems. 

To determine precisely the characteristics of the BBH systems
\begin{enumerate}
	\item we found the characteristics of the BBH system ejecting a component using 
	the coordinates of the component given by \citep{KuGa+:14},
	\item we estimated the perturbation due to the slow rotation of the BBH system 
	ejecting the component around the second 	BBH system assuming 
	$M_{BHC6} + M_{BH4} = (M_{CS} + M_{Cg}) /10$ and 	corrected the coordinates given 
	by \citep{KuGa+:14} from this perturbation, finally
	\item we used the corrected coordinates to find the final characteristics of the BBH system.
\end{enumerate}

The characteristics of the two BBH systems are
\begin{itemize}
	\item the distance between the two BBH systems is $\approx 1.35$ mas $\approx 6.0$ pc,
	\item the first BBH system is constituted by the components Cg and CS,
	\item the radius of the BBH system is $R_{bin,1} \approx 220$ $\mu$as $\approx 0.98$ pc, 
	\item the mass ratio of the two black holes is $M_{Cg} = 3 \times M_{CS}$,
	\item the coordinates of the two black holes are $X_{Cg} = 0$, $Y_{Cg} = 0$ 
	(by definition) and $X_{CS} \approx -70$ $\mu$as, $Y_{CS} \approx +210$ $\mu$as,
	\item the two black holes of the second BBH system, namely BHC6 and BH4 are 
	not associated with stationary VLBI components, i.e. they are not strong radio 
	sources, 
	\item the radius of the BBH system is $R_{bin,2} \approx 140$ $\mu$as $\approx 0.62$ pc, 
	\item the mass ratio of the two black holes is $M_{BHC6} =  0.12 \times M_{BH4}$ and
	\item the coordinates of the black hole BHC6 is 
	$X_{BHC6} \approx -100$ $\mu$as, $Y_{CS} \approx -1300$ $\mu$as and the coordinates 
	of BH4 are unknown.
\end{itemize}

We found that the inclination angle is between $i_{o} \approx 18.5^{\circ}$ and 
$i_{o} \approx 23.5^{\circ}$.

Combining the constraints obtained using the fits of components C1, C6 and C8, 
i.e. making the intersection of the ranges of the BBH systems parameters 
given in Tables 1, 2 and 3, we can 
deduce the characteristics of the BBH systems associated with the nucleus of 
1928+738. They are 

\begin{itemize}
	\item the total mass of the BBH system Cg-CS is $3 \; 10^{6}$ $M_{\odot}$ $\leq M_{Cg}+M_{CS} \leq 1.2 \: 10^{11}$ $M_{\odot}$,
	\item the period of the BBH system Cg-CS is $265$ yr $\leq T_{bin} \leq 52200$ yr,
	\item the size of the accretion disk around Cg is $R_{disk,Cg} \approx 0.027$ mas $\approx 0.12$ pc and the 
	rotation period of the disk is $13.2$ yr $\leq T_{disk,Cg} \leq 2590$ yr,
	\item the size of the accretion disk around CS is $R_{disk,CS} \approx 0.021$ mas $\approx 0.093$ pc and the 
	rotation period of the disk is $15.5$ yr $\leq T_{disk,Cg} \leq 3040$ yr,
	\item the total mass of the BBH system BHC6-BH4 is $3 \: 10^{5}$ $M_{\odot}$ $\leq M_{BHC6}+M_{BH4} \leq 1.2 \: 10^{10}$ $M_{\odot}$,
	\item the period of the BBH system BHC6-BH4 is $420$ yr $\leq T_{bin} \leq 84600$ yr,
	\item the size of the accretion disk around BHC6 is $R_{disk,Cg} \approx 0.006$ mas $\approx 0.027$ pc and the 
	rotation period of the disk is $12.4$ yr $\leq T_{disk,Cg} \leq 2500$ yr,
\end{itemize}
Combining the constraints obtained using the fits of components C1, C6 and C8 
reduce the range of the parameters obtained for each fit done separately.\\

\citet{RoKa+:93} assumed that the mass of the nucleus was $\approx 10^{8}$ $M_{\odot}$ 
and \citet{KeBe:07} estimated the mass of the nucleus to be $\approx 8 \: 10^{8}$ $M_{\odot}$. 
So, if we assume $M_{Cg}+M_{CS} \approx 8 \: 10^{8}$ $M_{\odot}$, we have 
$M_{Cg} \approx 6 \; 10^{8}$ $M_{\odot}$ and $M_{CS} \approx 2 \; 10^{8}$ $M_{\odot}$. 
We find that the orbital period of the BBH system (Cg-CS) is $T_{bin} \approx 3195$ yr, 
and the rotation periods of the accretion disks around Cg and CS are $T_{disk,Cg} \approx 159$ yr and 
$T_{disk,CS} \approx 186$ yr. Assuming $M_{BHC6}+M_{BH4} \approx (M_{Cg}+M_{CS})/10 \approx 8 \: 10^{7}$ $M_{\odot}$, 
we have $M_{BHC6} \approx 8.7 \; 10^{6}$ $M_{\odot}$ and $M_{BH4} \approx 7.1 \; 10^{7}$ $M_{\odot}$. 
We find that the orbital period of the BBH system (BHC6-BH4) is $T_{bin} \approx 5130$ yr, 
and the rotation period of the accretion disk around BHC6 is $T_{disk,Cg} \approx 152$ yr. 
We can also find the orbital period of the two black hole systems, 
with a mean distance $\approx 1.35$ mas, it is $T_{bin} \approx 46300$ yr. 
We can deduce the propagation speeds of the different families of 
trajectories. The propagation speed of the family corresponding the ejection of C8 is 
$V_{a,Cg} \approx 0.045$ c, 
the propagation speed of the family corresponding the ejection of C1 is $V_{a,CS} \approx 0.064$ c 
and the propagation speed of the family corresponding the ejection of C6 is $V_{a,BHC6} \approx 0.016$ c.

During the period of observations, i.e. about 20 years, the black holes associated 
with CS and Cg ejected quasi regularly VLBI components but the second BBH system ejected only 
three components within 8 years (Fig.~\ref{fig:6_1928+738_Birth_of_Components_2_Bis}). 
Note that there is no periodicity for the ejection of VLBI components.

\citet{KuGa+:14} found that the flux density of first two mas of the VLBI jet was 
quasi periodic with a period $\approx 4.5$ yr. The first two mas can contain VLBI 
components ejected by Cg and CS. The blue shift factor corresponding to component C1, 
ejected by CS, is $\approx 0.08$, and the blue shift factor corresponding to component C8, 
ejected by Cg, is $\approx 0.18$. The quasi period observed by \citet{KuGa+:14} was  
related to ejection of new VLBI components, and it 
corresponds, in the quasar frame, to a value between $\approx 25$ yr and $\approx 56$ yr. 
This period corresponds to a fraction, i.e. between 1/8 and 1/3  of the rotation periods 
of the accretion disks but does not correspond to the orbital period of the BBH system.


In the case of 1928+738, our modeling shows that the nucleus contains two 
BBH systems on the pc scale, i.e. the size of the binary systems 
are $R_{bin,1} \approx 0.22$ mas $\approx 0.98$ pc, 
$R_{bin,2} \approx 0.14$ mas $\approx 0.62$ pc, and the distance between the two 
BBH systems is $\approx 1.35$ mas $\approx 6$ pc. 
\citet{DePa+:14} reports the detection of triple system 
which size is $\approx 7.4$ kpc with a binary system which size is $\approx 140$ pc. 
The interpretation of \citet{DePa+:14} has been questionned by \citet{WoWa+:14}.   
For the formation of triple supermassive black hole systems, see \citet{HoLo:07} and 
\citet{KuLo:12}. As indicated in the Introduction, the ejection of VLBI components can 
be perturbed by the motion BBH system around a third black hole or an other BBH system. 
From VLBI observations there is a signature of this kind of perturbation, i.e. of the presence 
of a triple black hole system or a double BBH system in the nucleus. 
One observes close to the nucleus short period wiggles 
followed by a single turn which changes the ejection direction by a large angle, 
which can be 45 degrees. The best sources showing this behavior and which contains a 
triple system or two BBH systems are 3C 345 (work in progress) and 3C 454.3.\\


After the merging of two galaxies, if a BBH system form, it evolves 
\textit{rapidly} to reach a critical radius $R_{bin} \sim 1$ pc. 
After it loses energy emitting gravitational waves and it takes several billions 
years to collapse \citep{BrRo+:01}. This explains that the typical size of 
the BBH system found in nuclei of extragalactic radio sources is $0.25$ pc $\leq R_{bin} 
\leq 1.5$ pc \citep{Ro:14}. A typical BBH system which size is $\approx 1$ pc and 
which contain two similar black holes of $10^{8} M_{\odot}$ is characterized by 
an orbital period of 6600 yrs. The mean speed of the two black holes is $ \approx 950$ km/s. 
If the inclination angle of the source is $i_{o} \leq 10^{\circ}$, the difference 
between the radial speeds of the emission lines of the two cloud systems associated 
with the two black holes will be $\Delta V_{r} \leq 165$ km/s. This result shows:
\begin{enumerate}
	\item why it is difficult to detect BBH systems studying the broad lines spectra 
	of quasars, and
	\item the most efficient method to find BBH systems and to determine their 
	characteristics is to study the kinematics of ejected VLBI components.
\end{enumerate}


\begin{figure}[ht]
\centerline{
\includegraphics[scale=0.5, width=8cm,height=6cm]{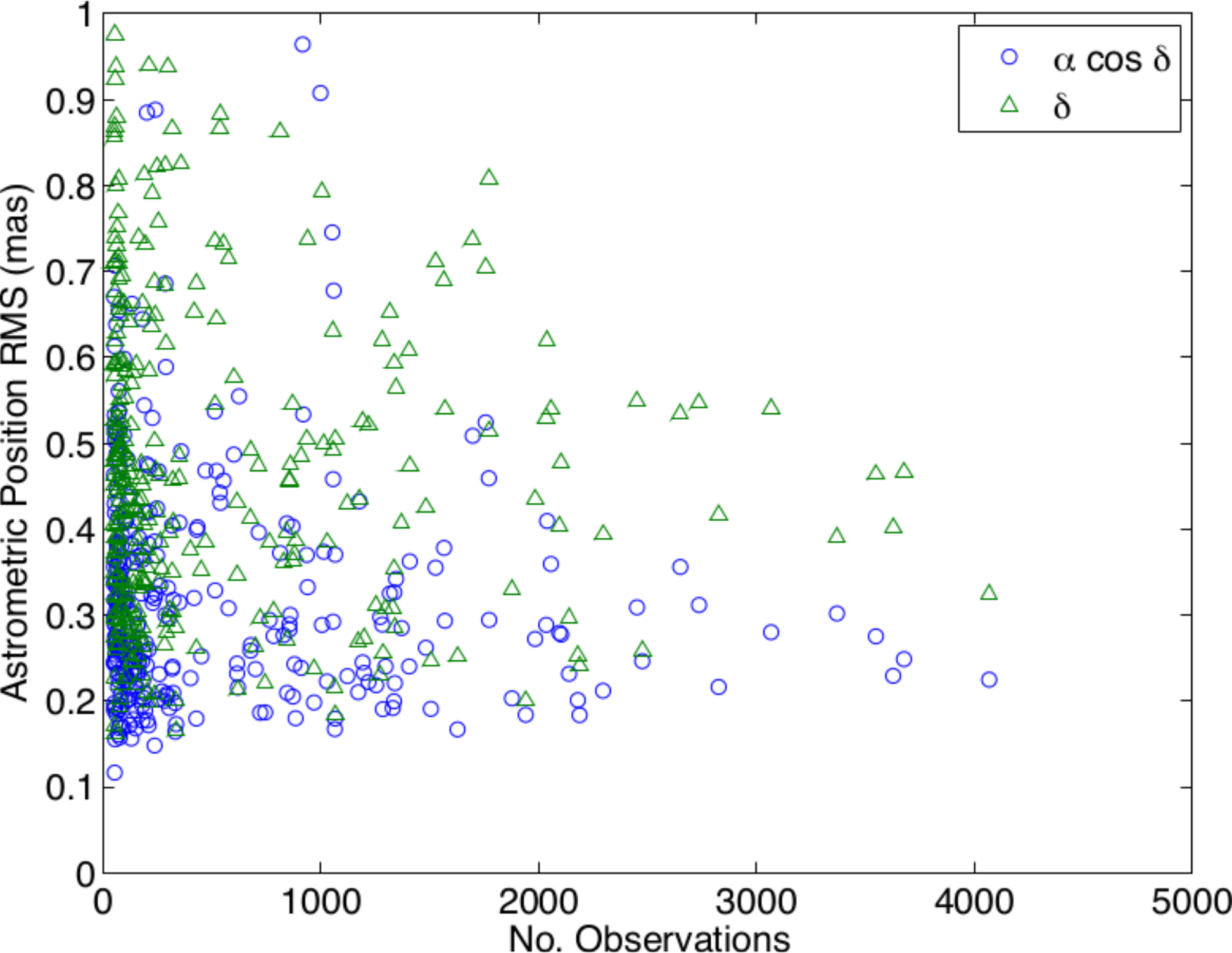}}
\caption{Rms of the coordinate time series of the most observed quasars 
in the geodetic VLBI monitoring program of the International VLBI Service 
for Geodesy and Astrometry (IVS) (Lambert 2014) as a function of the number of sessions.}
\label{fig:8_fig-series}
\end{figure}

\begin{figure}[ht]
\centerline{
\includegraphics[scale=0.5, width=8cm,height=6cm]{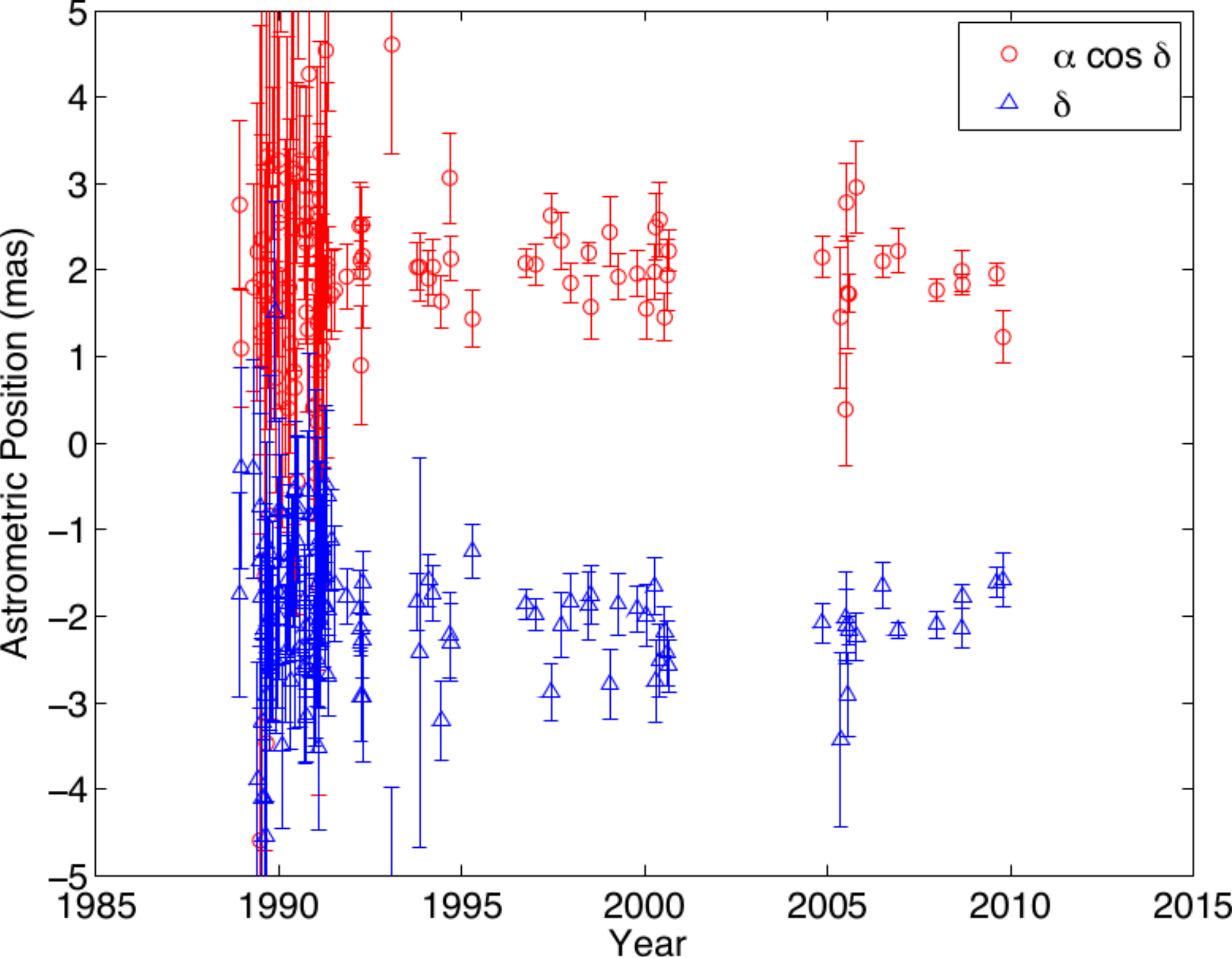}}
\caption{Coordinate time series (mean removed and shifted by 2 mas 
for $\alpha$ and $-2$ mas for $\delta$) of 1928+738 obtained by the 
analysis of data of the geodetic VLBI monitoring program of the IVS (Lambert 2014).}
\label{fig:9_fig-series2}
\end{figure}

As shown in Fig.~\ref{fig:8_fig-series}, the position of radio sources as 
measured by geodetic VLBI shows displacements larger than 0.1 mas in rms. 
This floor is obviously not only  due to changes in the radio source structure: 
several other limiting factors like the mis-modeling of the troposphere wet delay 
and the noise introduced by site-dependent correlated errors play a non negligible role. 
In the case of 1928+738, the size of the BBH system associated with Cg and Cs is 
0.22 mas and the distance between the two BBH systems is 1.35 mas. The corresponding 
coordinate time series (Fig.~\ref{fig:9_fig-series2}), computed after observations of the 
geodetic VLBI monitoring program of the International VLBI Service for Geodesy and 
Astrometry (IVS; \citet{ScBe:12}; \citet{La:14}) is close to 0.3 mas, which is 
in agreement with (i) the size of the former BBH system, and (ii) the fact that this 
BBH system is much more active than the latter. For geodetic VLBI observations, 
1928+738 therefore appears as a single BBH system of size 0.22~mas. However, the 
radio center detected by geodetic VLBI will follow the emitting black hole, and one 
can therefore expect significant displacements of the order of the size of the BBH system. 
Evidences of such a correspondence between the size of the BBH system, generally larger 
than 0.1 mas, and the rms of the coordinate time series has been raised in \citet{Ro:14} 
although this study considered only a very few sources and must be extended to other sources. 
If so, it is likely that the astrometric precision of VLBI will be limited in the future 
by the size of the BBH systems, even at frequencies higher than the current 8.6 GHz band 
used for the ICRF2. The determination of the number and sizes of BBH systems in quasar 
nuclei will therefore be crucial in the future realization of reference frames, especially 
for the choice of the so-called “defining sources” that define the system axes and that 
should be, in principle, as point-like as possible.

Same remarks can apply to the Gaia optical reference frame 
\citep{PeBo+:01}. 
If the nucleus of the radio quasar contains a BBH system and if the 
two black holes are active, three different cases can happen:
\begin{enumerate}
	\item the radio core and the optical core are associated with the same BH, then
the distance between the radio core and the optical core depends on the opacity
effect which will be small if the inclination angle is small,
  \item the radio core and the optical core are associated with different black holes, then
the distance between the radio core and the optical core is more or less the size
of the BBH system (corrected by the possible opacity effect), and
  \item the two black holes are emitting in the optical, then
GAIA will provide a mean position between the two optical cores. 
This position will be different from the positions of the two radio cores.
\end{enumerate}

As quasars are strongly and rapidly variables, during the 5 years of 
observations of GAIA, the 3 different cases can happen for a given source.

Although that in the case of 1928+738, it is associated with a bright optical quasar 
and the radio positions of the 2 radio emitting black holes are known, 
\begin{itemize}
	\item the large value of the inclination angle implies that opacity effect 
	can be significant and
	\item the fact that the two black holes are active and ejecting VLBI components,
\end{itemize}
make the use of 1928+738 to obtain the precise link between the radio positions 
and the optical position, obtained by GAIA, difficult.

\begin{acknowledgements}
We thank the referee for very useful comments and 
one of us (JR) thanks Bertha Sese for enlightening discussions. 
This research has made use of data from the MOJAVE 
database that is maintained by the MOJAVE team \citep{LiAl+:09} and part of 
this work was supported by the COST Action MP0905 Black Holes in a Violent Universe. 

\end{acknowledgements}

\bibliographystyle{aa} 
\bibliography{JRoland_BIB} 

\newpage
\appendix
\section{Fit of component C8}
\label{Fit_C8_1928+738}

\subsection{Introduction}
The trajectory of C8 is observed 
for the first 5 mas, however the map of 1928+738 (Fig.~\ref{fig:1_1928+738_2cm03_08_Bis_3}) 
shows that the VLBI jet turns after about 10 mas \citep{LiHo:05}.

In a first step, we will try to find a solution of which could explain the long term 
turn using a BBH system, in a second step we will study the influence of a third black hole 
or a second BBH system on the solution and correct the coordinates of \citet{KuGa+:14} 
from this perturbation and finally determine the characteristics of the BBH 
system ejecting C8 using the corrected coordinates.

We used the method developed by \citet{RoBr+:08} and \citet{RoBr+:13}.

\subsection{Solution using the coordinates of \citet{KuGa+:14}}

The solution obtained corresponds to a VLBI ejection which asymptotic direction 
is $\Delta \Xi \approx 182\degr$.\\

The main characteristics of the BBH system ejecting C8 are that
\begin{itemize}
    \item the two black holes are associated with the components CS and Cg,
		\item the VLBI component C8 is ejected by the VLBI core, i.e. component Cg 
		(there is no indication of an offset of the origin of the ejection), 
    \item the radius of the BBH system is $R_{bin} \approx 220$ $\mu as$ $\approx 0.98$ $pc$,
    \item the ratio $M_{Cg}/M_{CS}$ is $\approx 4$, which is the inverse of the mass ratio found 
		fitting the coordinates of C1 (Sec.~\ref{Fit_C1_1928+738})  and
		\item the ratio $T_{p}/T_{b}$ is $\approx 129$ .
\end{itemize}

\begin{figure}[ht]
\centerline{
\includegraphics[scale=0.5, width=8cm,height=6cm]{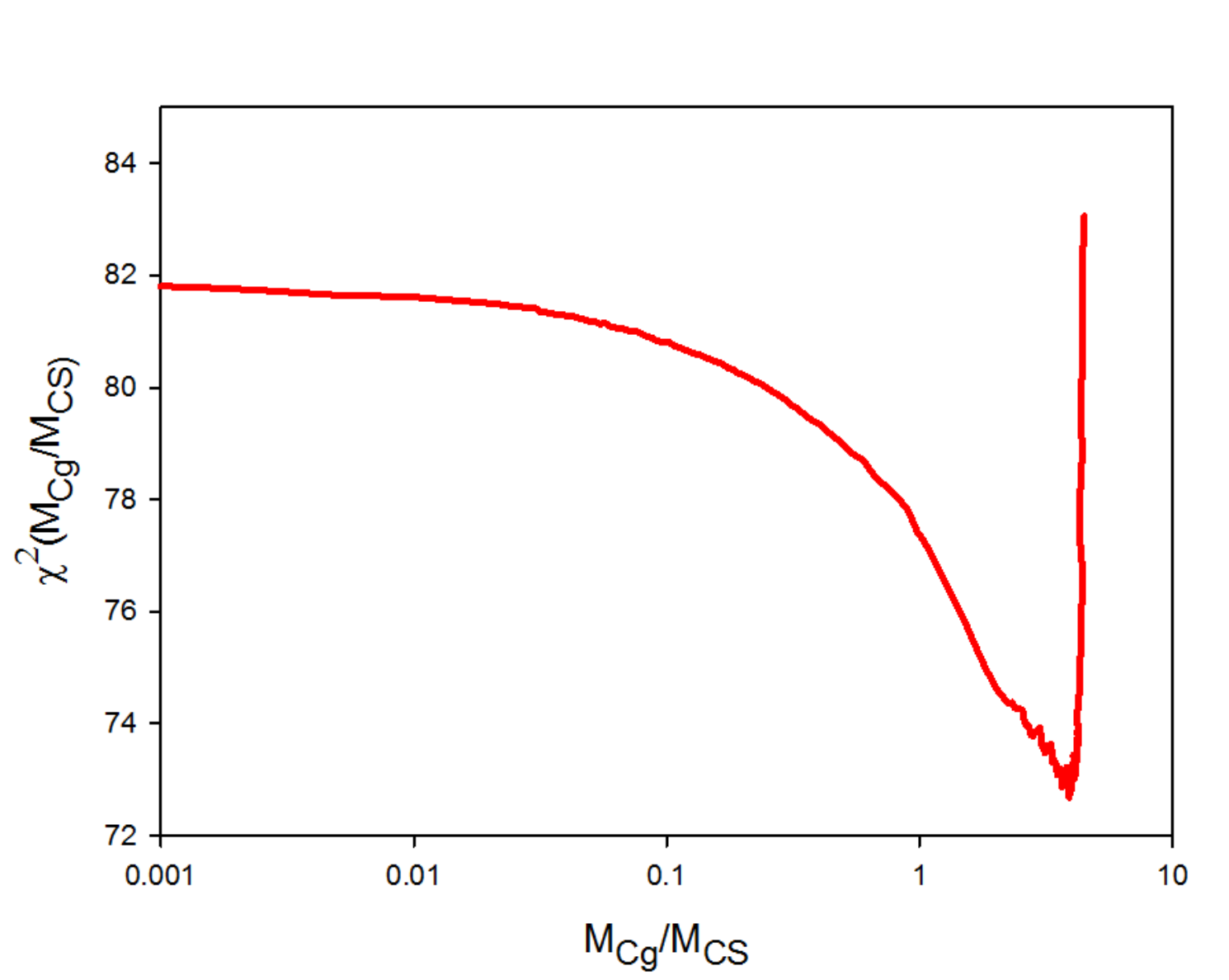}}
\caption{Determination of the parameter $M_{Cg}/M_{CS}$. 
We calculated $\chi^{2}(M_{Cg}/M_{CS})$ fitting of coordinates of C8, this provides the value 
of the ratio $M_{Cg} /M_{CS}$, i.e. $M_{Cg} /M_{CS} \approx 4$.}
\label{fig:10a_Chi2_Par_io20_Rbin220_TpTb320_Final}
\end{figure}

The ratio $M_{Cg}/M_{CS}$ is a free parameter of the model and the value $M_{Cg}/M_{CS} \approx 4$ 
comes from the fit of the coordinates of C8 (see Fig.~\ref{fig:10a_Chi2_Par_io20_Rbin220_TpTb320_Final}).\\

We find also that
\begin{itemize}
    \item the inclination angle is $i_{o} \approx 26^{\circ}$,
    \item the angle between the accretion disk and the rotation plane 
    of the BBH system is $\Omega \approx 9.2^{\circ}$,
    \item the bulk Lorentz factor of the VLBI component is $\gamma_{c} \approx 5.5$, and
    \item the origin of the ejection of the VLBI component is $t_{o} \approx 1993.9$.
\end{itemize}

\begin{figure}[ht]
\centerline{
\includegraphics[scale=0.5, width=8cm,height=12cm]{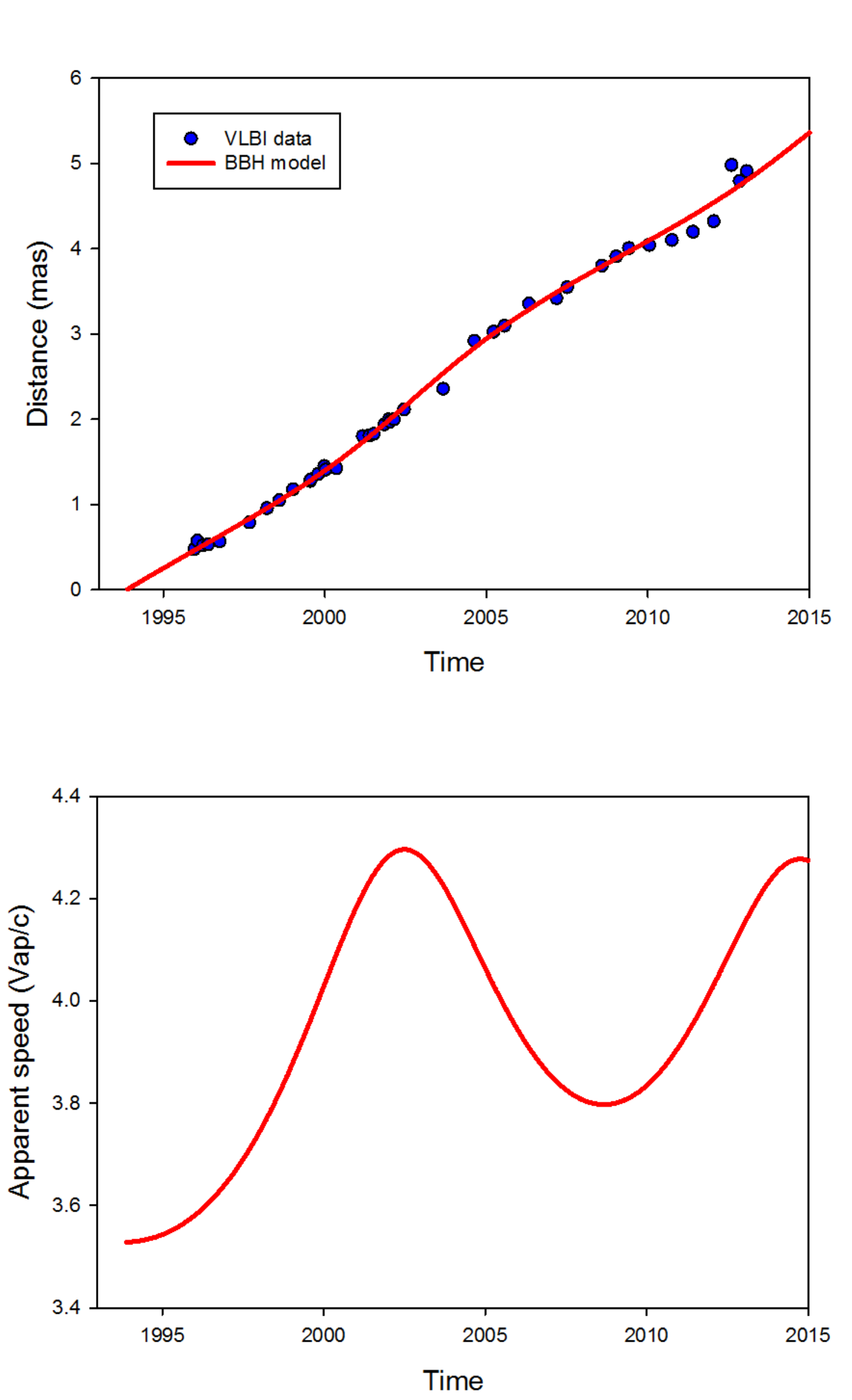}}
\caption{Variations of the distance and the apparent speed of component C8 
assuming a constant bulk Lorentz factor $\gamma_{c} \approx 5.5$. 
\textit{Top Figure}: From the plot of the variations of the distance we can deduce the mean speed: 
$4.86$ mas $/19.7$ yr $\approx 247$ $\mu$as /yr.
\textit{Bottom Figure}: Although the large value of the inclination angle, we 
observe superluminal motion with a mean speed $\approx 4.1$ c.}
\label{fig:10_Comp8_Distance+Vap_Time_15G_New4}
\end{figure}

The variations of the distance and the apparent speed of component C8 are showed 
in Fig.~\ref{fig:10_Comp8_Distance+Vap_Time_15G_New4}. We find that component C8 moves with a 
mean apparent speed $v_{ap} \approx 4.1$ c, a value similar to that obtained by \citet{LiAl+:13}.

\begin{figure}[ht]
\centerline{
\includegraphics[scale=0.5, width=8cm,height=12cm]{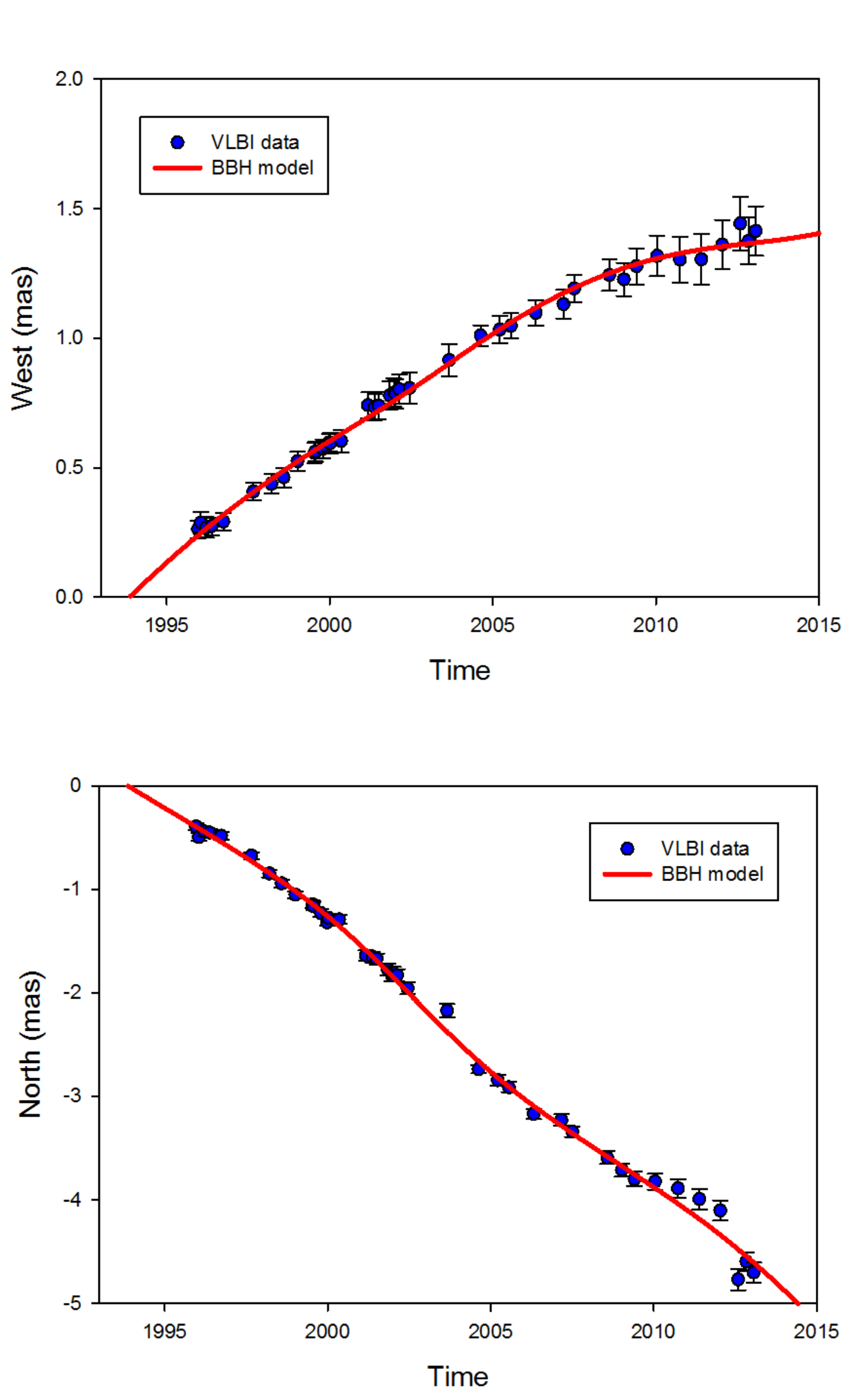}}
\caption{Fits of the two coordinates $W(t)$ and $N(t)$ of component C8 of 1928+738. 
They correspond to the solution with $T_{p}/T_{b} \approx 130$, 
$M_{Cg}/ M_{CS} \approx 4$, and $i_{o} \approx 26^{\circ}$. The points are the observed 
coordinates of component C8. The VLBI coordinates and their error bars are taken from \citet{KuGa+:14}. 
The red lines are the coordinates of the component trajectory calculated using the BBH model.}
\label{fig:11_Comp8_Xt+Yt_15G_New4}
\end{figure}

The fits of the two coordinates $W(t)$ and $N(t)$ of the component C8 of 1928+738 are  
showed in Fig.~\ref{fig:11_Comp8_Xt+Yt_15G_New4}.

\subsection{Determining the family of solutions}
For the inclination angle previously found, i.e., 
$i_{o}\approx 26^{\circ}$, $T_{p}/T_{b} \approx 129$, $M_{Cg}/M_{CS} \approx 4$, 
and $R_{bin} \approx 220$ $\mu as$, we gradually varied $V_{a}$ between $0.001$ c 
and $0.45$ c. The function $\chi^{2}(V_{a})$ remained constant,  
indicating a degeneracy of the solution. We deduced the 
range of variation of the BBH system parameters. They are given in Table 4.

\begin{center}
Table 4 : Ranges for the BBH system parameters ejecting C8\medskip%

\begin{tabular}
[c]{c||c|c}\hline
$V_{a}$                    & $0.001 \: c$                            & $0.45 \: c$                            \\\hline
$T_{p}(V_{a})$             & $\approx 12000000$ yr                   & $\approx 14000$ yr                     \\\hline
$T_{b}(V_{a})$             & $\approx 92000$ yr                      & $\approx 108$ yr                       \\\hline
$(M_{Cg}+ M_{CS})(V_{a})$  & $\approx 9.7 \; 10^{5}$ $M_{\odot}$     & $\approx 7.0 \; 10^{11}$ $M_{\odot}$   \\\hline
\end{tabular}
\end{center}

Table 4 provides the range of the BBH system parameters ejecting C8. To obtain the final range of the 
two BBH systems Cg-CS and BHC6-BH4 one has to make the intersection of the ranges of BBH systems parameters 
found after the fits of C8, C1 and C6 (see Sec.~\ref{section:Discussion_conclusion}).

For $V_{a,Cg} = 0.1$ c, we find that the total mass of the BBH system ejecting C8 is 
$M_{Cg} + M_{CS} \approx 1.26 \times 10^{10}$ $M_{\odot}$.

\subsection{Determining the size of the accretion disk}

From the knowledge of the mass ratio $M_{Cg}/M_{CS} \approx 4$ and 
the ratio $T_{p}/T_{b} \approx 130$, we calculated in the previous section 
the mass of the ejecting black hole $M_{Cg}$, the orbital period $T_{b}$, 
and the precession period $T_{p}$ for each value of $V_{a}$.

The rotation period of the accretion disk, $T_{disk}$, is given by 
(\ref{eq:Tdisk}). Thus we calculated the rotation period of the accretion disk, and 
assuming that the mass of the accretion disk is $M_{disk} \ll M_{Cg}$, 
the size of the accretion disk is given by (\ref{eq:Rdisk}). 
We found that the size of the accretion disk does not depend on $V_{a}$ and is 
$R_{disk}  \approx 0.028 \; mas \approx 0.124 \; pc $.

\section{Circular orbit correction of C8 coordinates}
\label{COC_C8_1928+738}

We found that the long term turn of the VLBI trajectory at about 10 mas could be 
explained by the BBH system associated with CS and Cg. However, components C5 
and C6 are ejected by either a third black hole or a second BBH system, so we have 
to estimate the influence the slow rotation of the BBH system Cg-CS around the mass 
ejecting C5 and C6 and correct the coordinates of C8 from this perturbation to 
make a new determination of the characteristics of the BBH system ejecting C8.

Let us call $M_{BH3}$ the mass ejecting C6, as the long term turn can be partly 
explained by the BBH system Cg-CS we should have $M_{BH3} /(M_{Cg}+M_{CS}) < 1$. 
We calculated the circular orbit correction for $M_{BH3} /(M_{Cg}+M_{CS}) = 1/10$, $1$ and $10$ 
and we found that we can find a non mirage solution only if $M_{BH3} /(M_{Cg}+M_{CS}) < 1$. 
So to continue, we chosen arbitrarily the ratio $M_{BH3} /(M_{Cg}+M_{CS}) = 1/10$. 
In order to estimate the influence of the choice of this value on the 
final numerical result one could calculate the circular orbit correction assuming 
for instance $M_{BH3} /(M_{Cg}+M_{CS}) = 1/5$. This choice $M_{BH3} /(M_{Cg}+M_{CS}) = 1/5$ 
will not change the conclusion but simply the numerical result. 

Using the parameters of the solution found in Sec.~\ref{Fit_C8_1928+738}, i.e. for 
$V_{a} = 0.1$ c, we have $M_{Cg}+M_{CS} \approx 1.26 \times 10^{10}$ $M_{\odot}$ and then 
$M_{BH3} \approx 1.26 \times 10^{9}$ $M_{\odot}$. As the distance between the BBH system 
Cg-CS and BH3 is $\approx 1.35$ $mas$ (see Sec.~\ref{Fit_C6_1928+738}), 
the corresponding orbital period of rotation of Cg-CS around BH3 is $T_{bin} \approx 11671$ yr. 
Keeping the geometrical parameters of the solution found in Sec.~\ref{Fit_C8_1928+738} 
we calculate the trajectory and the tangent to the trajectory. At a given time, knowing 
the coordinates, $W_{CO}(t)$, $N_{CO}(t)$, of the trajectory of the VLBI component due 
to the slow circular orbit motion, and the coordinates, $W_{tan}(t)$, $N_{tan}(t)$, 
of the VLBI component along the tangent trajectory, the VLBI coordinates corrected from the slow 
orbital motion are 

\begin{equation}
	W_{cor}(t) = W(t) - (W_{CO}(t) - W_{tan}(t)),
	\label{eq:W_cor}
\end{equation}
\begin{equation}
	N_{cor}(t)$ = $N(t) - (N_{CO}(t) - N_{tan}(t)).
	\label{eq:N_cor}
\end{equation}

\begin{figure}[ht]
\centerline{
\includegraphics[scale=0.5, width=8cm,height=6cm]{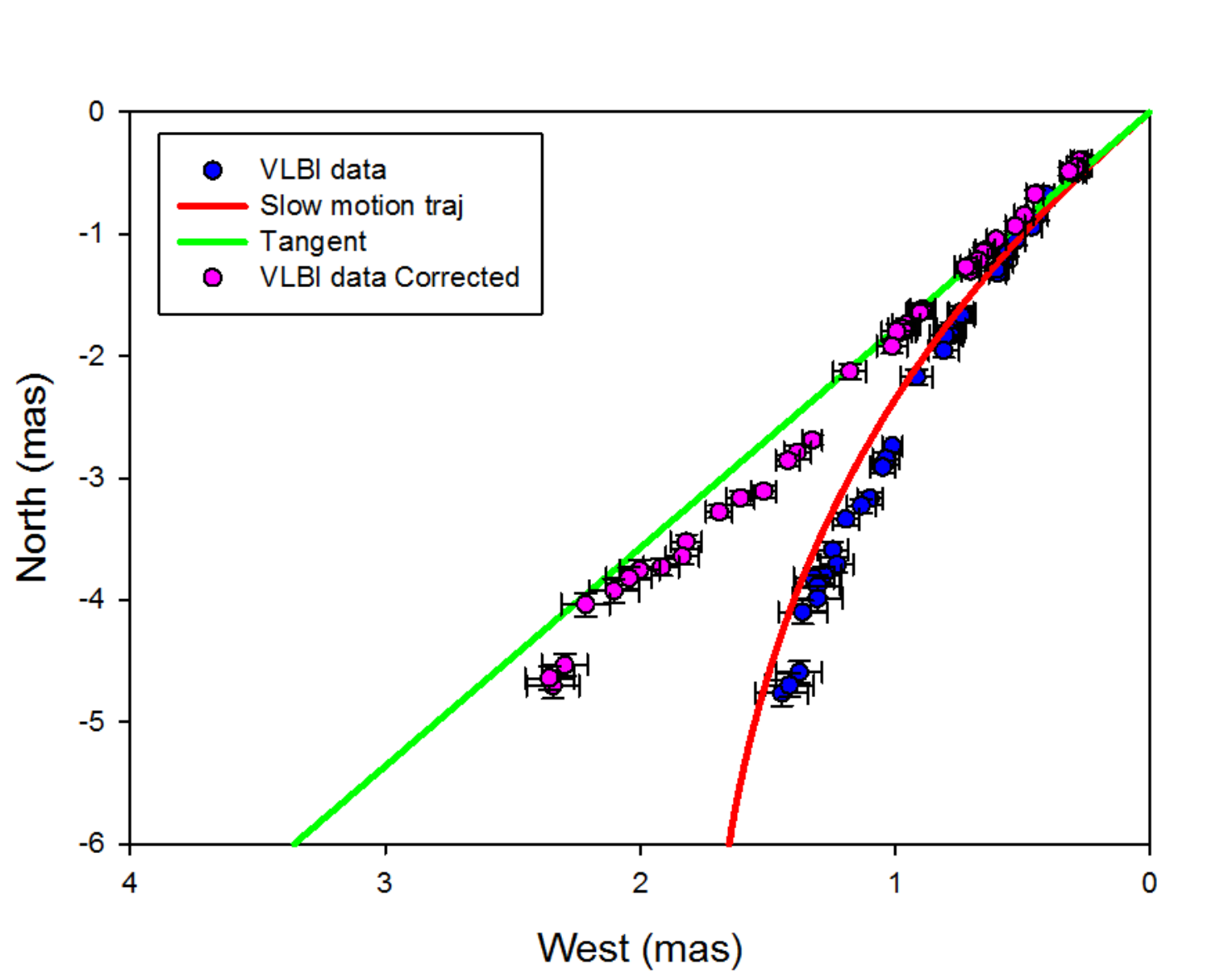}}
\caption{Plot of the trajectory of the VLBI component due to the slow circular orbit motion, 
the tangent trajectory, the VLBI coordinates given by \citet{KuGa+:14} and the coordinates 
corrected from the slow orbital motion.}
\label{fig:12_C8M_Traj_15G_New4_COC_BH3_182_10_-+_Bis}
\end{figure}

We plotted in Fig.~\ref{fig:12_C8M_Traj_15G_New4_COC_BH3_182_10_-+_Bis}, the trajectory of the 
VLBI component due to the slow circular orbit motion, the tangent trajectory, the VLBI coordinates 
given by \citet{KuGa+:14} and the coordinates corrected from the slow orbital motion.

Using the corrected VLBI coordinates, we made a new determination of the characteristics 
of the BBH system ejecting component C8. The result is given in Sec.~\ref{sec:solution_C8_1928+738}

\section{Fit of component C7a}
\label{Fit_C7a_1928+738}
We assumed that component C7a belongs to the family 
of components ejected by the black hole Cg. 
To check this hypothesis and the consistency of the model found, we will use the characteristics 
of the BBH system Cg-CS and the characteristics of the geometrical parameters of the trajectory 
of C8, to fit the coordinates of components C7a.

If C7a has been ejected by Cg, we have to fit the coordinates of C7a using 
the characteristics of the BBH system Cg-CS found in Sec.~\ref{Fit_C8_1928+738}, i.e.
\begin{itemize}
  \item Cg is the origin of the ejection,
	\item $T_{p} \approx 103998$ yr,
	\item $T_{p} / T_{b} \approx 129$,
	\item $R_{bin} \approx 0.220$ $\mu$as and 
	\item $M_{Cg} /M_{CS} \approx 4$,
\end{itemize}
and using the same geometrical parameters than those found to fit the trajectory of C8, i.e. 
\begin{itemize}
	\item $\Delta\Xi \approx 182^{\circ}$,
	\item $\Omega \approx 9.2^{\circ}$,
	\item $R_{o} \approx 28$ pc and  
	\item $T_{d}\approx 710$ yr.
\end{itemize}

Then we calculate $\chi^{2}(i_{o})$ starting from $i_{o} \approx 26^{\circ}$ 
and assuming that the parameters:
\begin{itemize}
	\item $\phi_{o}$ the phase of the precession at $t_{o}$,
	\item $\gamma_{c}$ the bulk Lorentz,
	\item $\Psi_{o}$ the phase of the BBH system at $t_{o}$ and 
	\item $t_{o}$ the time origin of the ejection of the component,
\end{itemize}
are free parameters.

\begin{figure}[ht]
\centerline{
\includegraphics[scale=0.5, width=8cm,height=6cm]{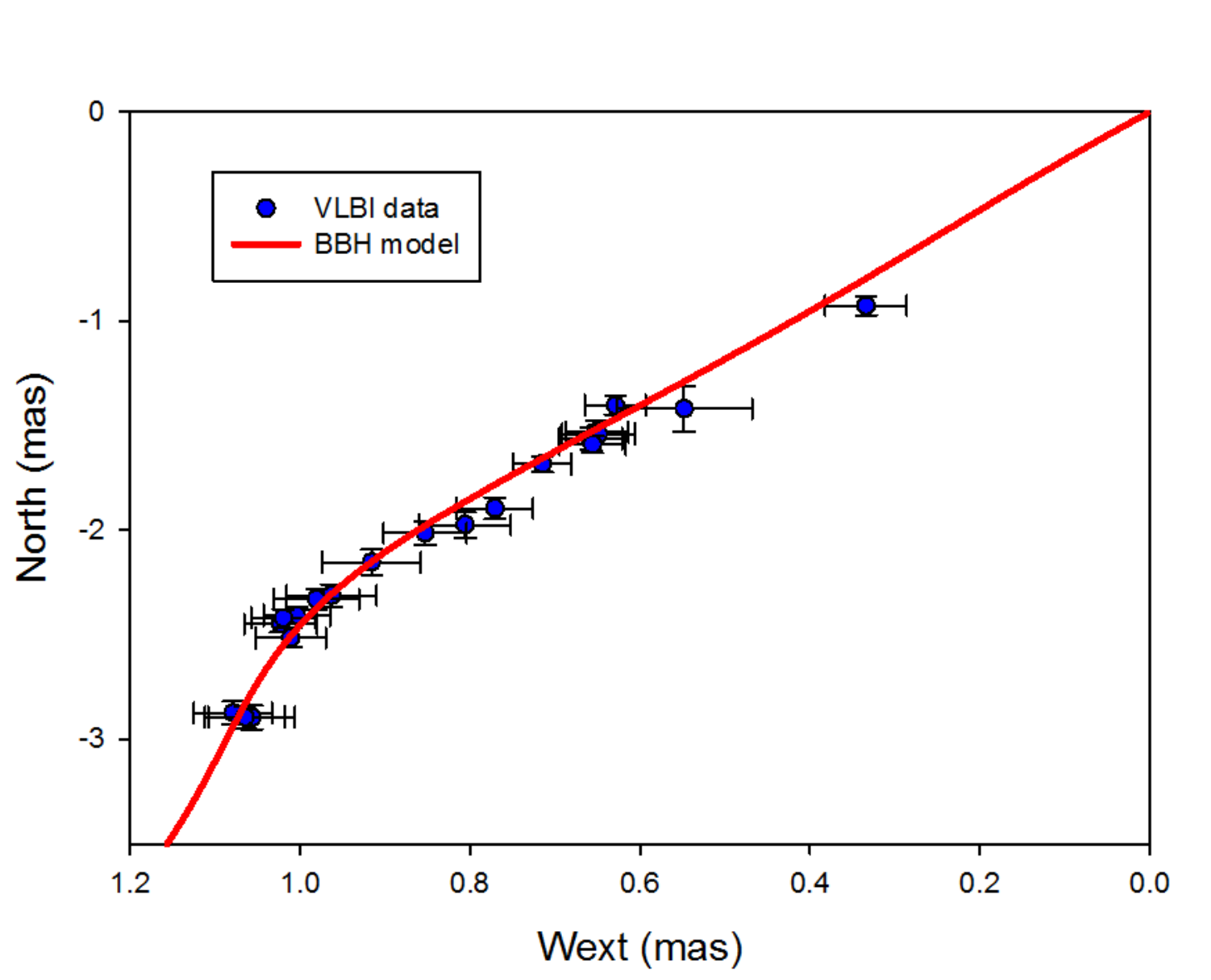}}
\caption{Trajectory of C7a assuming that it has been ejected by the black hole 
Cg of the BBH system Cg-CS and using the characteristics of the BBH system Cg-CS obtained 
during the fit of component C8 and the geometrical parameters of the trajectory of C8.}
\label{fig:12a_comp7a_v4_Traj_New1}
\end{figure}

The best fit is obtained for $i_{o} \approx 20^{\circ}$. The bulk Lorentz factor is 
$\gamma \approx 4.1$ and the time origin of the ejection is $t_{o} \approx 1992$. 
The trajectory of C7a is shown in Fig.~\ref{fig:12a_comp7a_v4_Traj_New1}. 
We obtain a very good fit of each coordinate showing that 
\begin{itemize}
	\item component C7a has been ejected by Cg, 
	\item the characteristics of the BBH system Cg-CS are correct and
	\item the solution found for the ejection of component C8 is the correct one.
\end{itemize}

\section{Fit of component C1}
\label{Fit_C1_1928+738}

The solution obtained corresponds to a VLBI ejection which asymptotic direction 
is $\Delta \Xi \approx 182\degr$.

The main characteristics of the solution of the BBH system 
associated with 1928+738 are that
\begin{itemize}
    \item the VLBI component C1 is not ejected by the VLBI core Cg, but by component CS 
		(there is a weak indication of an offset of the origin of the ejection in 
		the direction of CS, this weak indication is due to the lack of observations of C1 
		for the beginning of the trajectory),
		\item the coordinates of CS are $X_{CS} \approx -0.07$ mas and $Y_{CS} \approx +0.21$ mas,
    \item the radius of the BBH system is $R_{bin} \approx 220$ $\mu as$ $\approx 0.98$ $pc$,
    \item the ratio $M_{CS}/M_{Cg}$ is $\approx 1/4$, which is the inverse of the mass ratio found 
		fitting the coordinates of C8 (Sec.~\ref{Fit_C8_1928+738}) and
		\item the ratio $T_{p}/T_{b}$ is $\approx 53$ .
\end{itemize}

\begin{figure}[ht]
\centerline{
\includegraphics[scale=0.5, width=8cm,height=6cm]{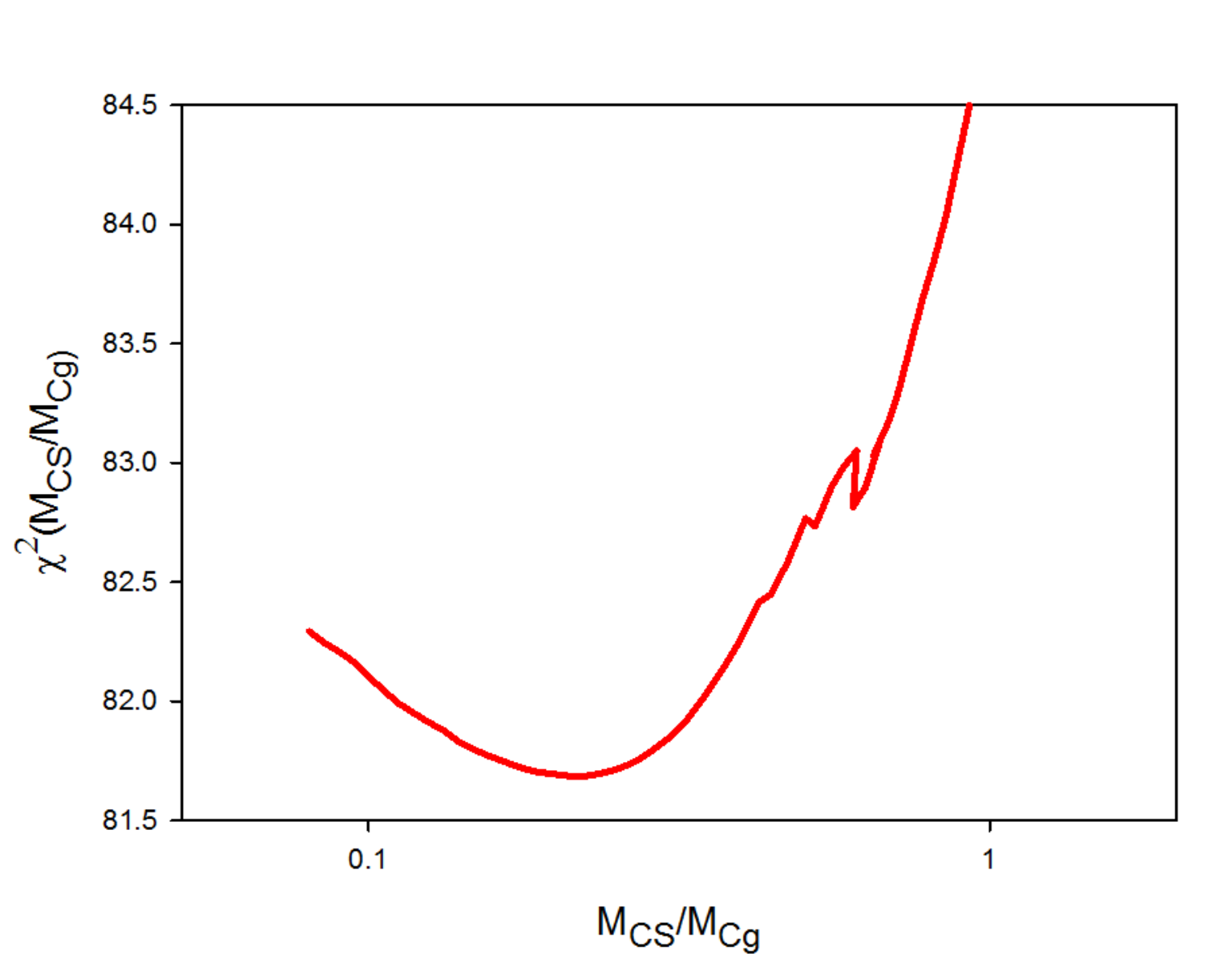}}
\caption{Determination of the parameter $M_{CS}/M_{Cg}$. 
We calculated $\chi^{2}(M_{CS}/M_{Cg})$ fitting of coordinates of C1, this provides the value 
of the ratio $M_{CS} /M_{Cg}$, i.e. $M_{CS} /M_{Cg} \approx 0.25$.}
\label{fig:13a_Chi2_par_io14_Rb220_182_TpTb320_Final}
\end{figure}

The ratio $M_{CS}/M_{Cg}$ is a free parameter of the model and the value $M_{CS}/M_{Cg} \approx 0.25$ 
comes from the fit of the coordinates of C1 (see Fig.~\ref{fig:13a_Chi2_par_io14_Rb220_182_TpTb320_Final}). 
The fact that the fit of C1 provides a mass ratio $M_{CS}/M_{Cg} \approx 0.25$ which is the inverse 
of the mass ratio $M_{Cg}/M_{CS} \approx 4$ obtained from the fit of component C8 shows that components 
C1 and C8 are not ejected by the same black hole.\\

We find that
\begin{itemize}
    \item the inclination angle is $i_{o} \approx 23^{\circ}$,
    \item the angle between the accretion disk and the rotation plane 
    of the BBH system is $\Omega \approx 10.3^{\circ}$,
    \item the bulk Lorentz factor of the VLBI component is $\gamma_{c} \approx 5.9$, and
    \item the origin of the ejection of the VLBI component is $t_{o} \approx 1967$.
\end{itemize}

\begin{figure}[ht]
\centerline{
\includegraphics[scale=0.5, width=8cm,height=12cm]{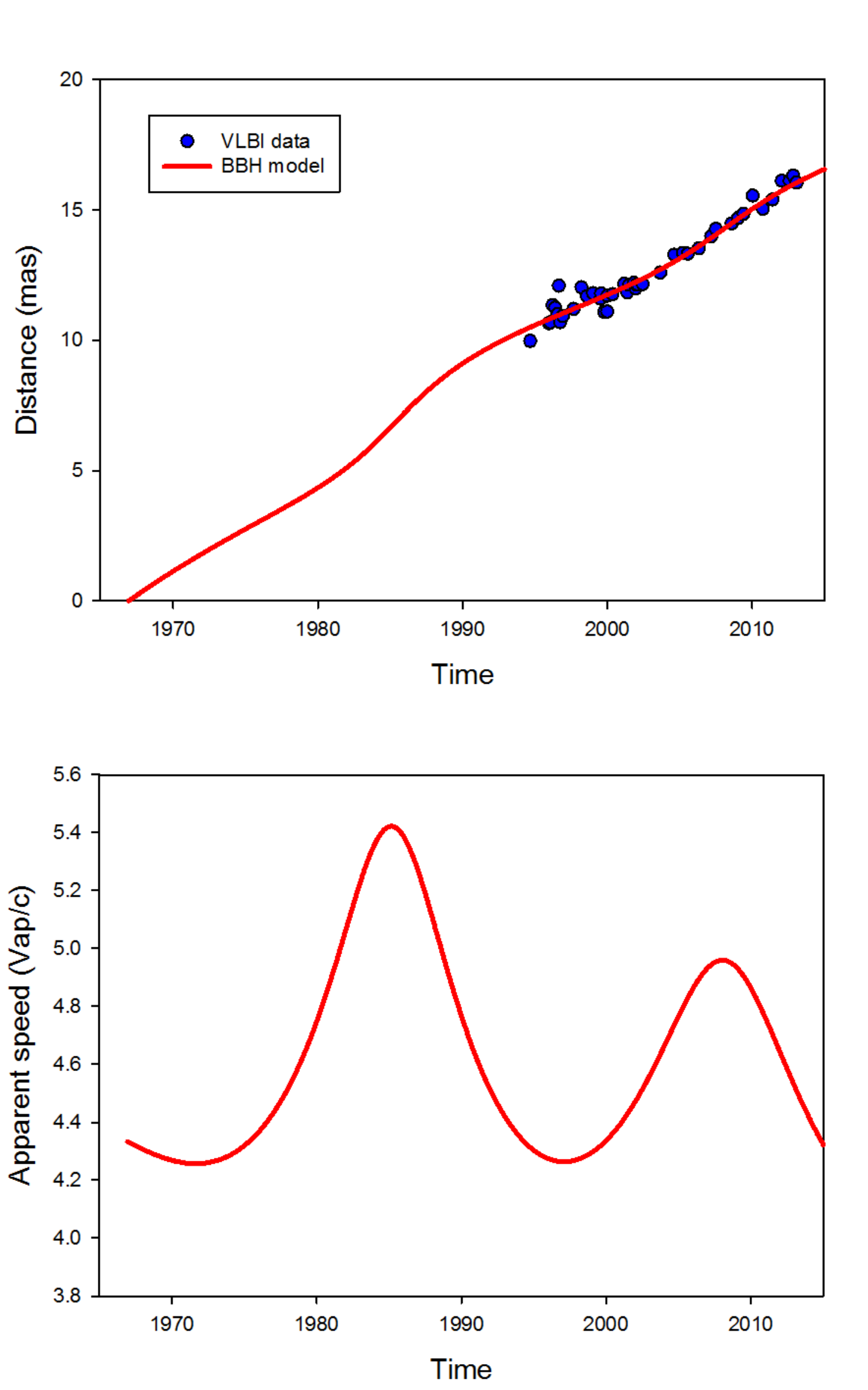}}
\caption{Variations of the distance and the apparent speed of component C1 
assuming a constant bulk Lorentz factor $\gamma_{c} \approx 5.9$. 
\textit{Top Figure}: From the plot of the variations of the distance we can deduce the mean speed: 
$16$ mas $/46$ yr $\approx 350$ $\mu$as /yr.
\textit{Bottom Figure}: Although the large value of the inclination angle, we 
observe superluminal motion with a mean speed $\approx 4.7$ c.}
\label{fig:13_SOLUTION_C1_Distance+Vap_Time_Off}
\end{figure}

The variations of the distance and the apparent speed of component C1 are showed 
in Fig.~\ref{fig:13_SOLUTION_C1_Distance+Vap_Time_Off}. We find that component C1 moves with a 
mean apparent speed $v_{ap} \approx 4.7$c, a value smaller that the one obtained by \citet{LiAl+:13}.

\begin{figure}[ht]
\centerline{
\includegraphics[scale=0.5, width=8cm,height=12cm]{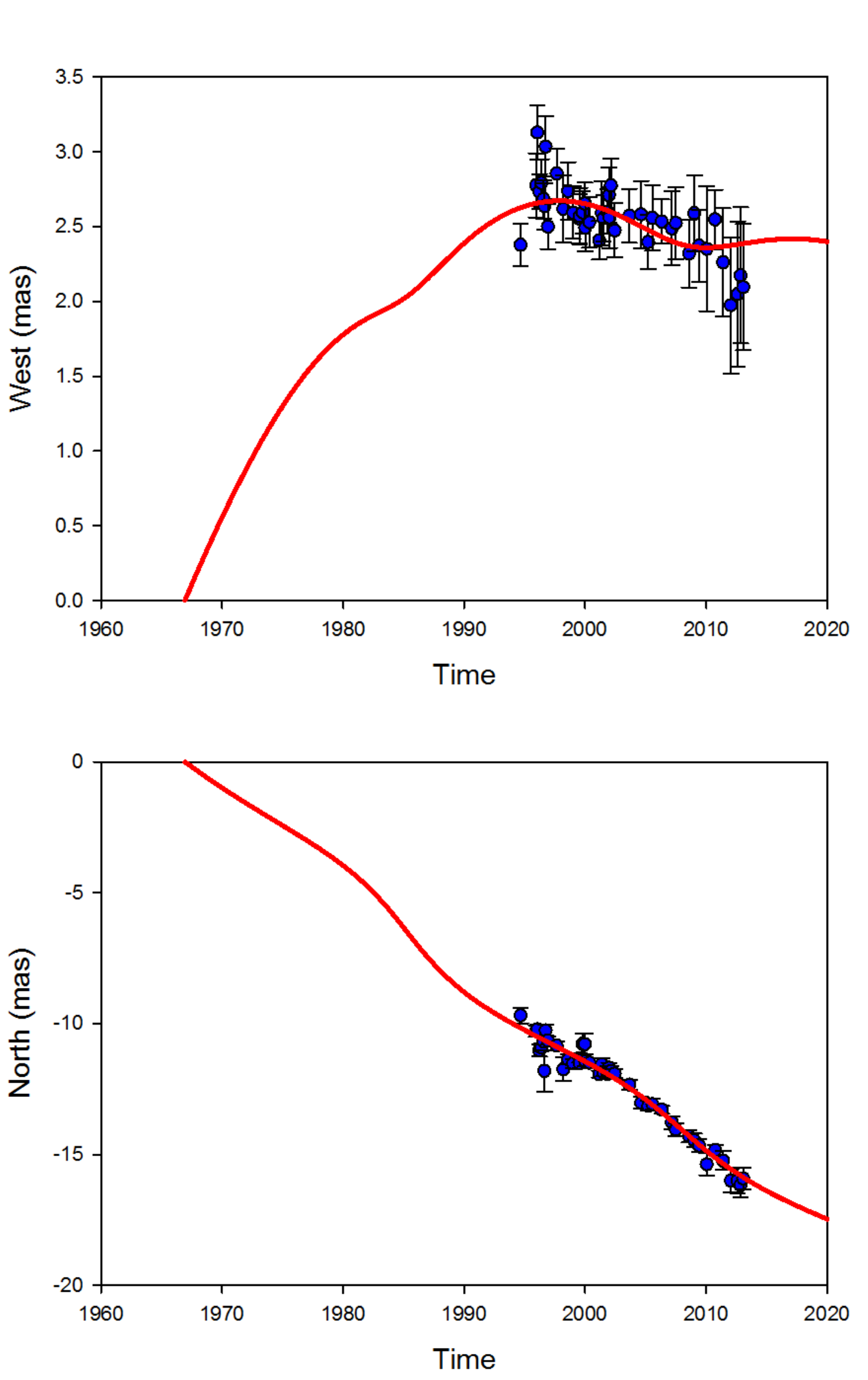}}
\caption{Fit of the two coordinates $W(t)$ and $N(t)$ of component C1 of 1928+738. 
They correspond to the solution with $T_{p}/T_{b} \approx 53$, 
$M_{CS}/ M_{Cg} \approx 1/4$, and $i_{o} \approx 23^{\circ}$. The points are the observed 
coordinates of component C1 that were corrected by the offsets $\Delta W \approx -70$ $\mu as$ 
and $\Delta N \approx +210$ $\mu as$. The VLBI coordinates and their error bars 
are taken from \citet{KuGa+:14}. The red lines are the coordinates of the 
component trajectory calculated using the BBH model.}
\label{fig:14_SOLUTION_C1_Xt+Yt_Time_Off}
\end{figure}

The fit of the two coordinates $W(t)$ and $N(t)$ of the component C1 of 1928+738 is  
showed in Fig.~\ref{fig:14_SOLUTION_C1_Xt+Yt_Time_Off}. 
The points are the observed coordinates of component C1 that were corrected 
by the offsets $\Delta W \approx -70$ $\mu as$ and $\Delta N \approx +210$ $\mu as$, and 
the red lines are the coordinates of the component trajectory calculated using the BBH model.

\subsection{Determining the family of solutions}
For the inclination angle previously found, i.e., 
$i_{o}\approx 23^{\circ}$, $T_{p}/T_{b} \approx 53$, $M_{CS}/M_{Cg} \approx 1/4$, 
and $R_{bin} \approx 220$ $\mu as$, we gradually varied $V_{a}$ between $0.001$ c 
and $0.45$ c. The function $\chi^{2}(V_{a})$ remained constant,  
indicating a degeneracy of the solution. We deduced the 
range of variation of the BBH system parameters. They are given in Table 5.

\begin{center}
Table 5 : Ranges for the BBH system parameters ejecting C1\medskip%

\begin{tabular}
[c]{c||c|c}\hline
$V_{a}$                    & $0.001 \: c$                          & $0.45 \: c$                            \\\hline
$T_{p}(V_{a})$             & $\approx 12400000$ yr                 & $\approx 14900$ yr                     \\\hline
$T_{b}(V_{a})$             & $\approx 230000$ yr                   & $\approx 279$ yr                       \\\hline
$(M_{Cg}+ M_{CS})(V_{a})$  & $\approx 1.5 \; 10^{5}$ $M_{\odot}$   & $\approx 1.0 \; 10^{11}$ $M_{\odot}$   \\\hline
\end{tabular}
\end{center}

Table 5 provides the range of the BBH system parameters ejecting C1. To obtain the final range of the 
two BBH systems Cg-CS and BHC6-BH4 one has to make the intersection of the ranges of BBH systems parameters 
found after the fits of C8, C1 and C6 (see Sec.~\ref{section:Discussion_conclusion}).

For $V_{a,CS} = 0.1$ c, we find that the total mass of the BBH system ejecting C1 is 
$M_{Cg} + M_{CS} \approx 1.87 \times 10^{9}$ $M_{\odot}$. The total mass of the BBH system 
ejecting C1 is the same that the total mass of the BBH system ejecting C8 if $V_{a,CS} \approx 0.232$ c, 
i.e. the propagation speeds of the perturbations are different for different families of trajectories 
(see Sec.~\ref{section:Discussion_conclusion} for the determination of the propagation speeds of the 
perturbations of three families of trajectories if $M_{Cg} + M_{CS} \approx 8 \times 10^{8}$ $M_{\odot}$).

\subsection{Determining the size of the accretion disk}

From the knowledge of the mass ratio $M_{Cs}/M_{Cg} \approx 1/4$ and 
the ratio $T_{p}/T_{b} \approx 53$, we calculated in the previous section 
the mass of the ejecting black hole $M_{CS}$, the orbital period $T_{b}$, 
and the precession period $T_{p}$ for each value of $V_{a}$.

The rotation period of the accretion disk, $T_{disk}$, is given by 
(\ref{eq:Tdisk}). Thus we calculated the rotation period of the accretion disk, and 
assuming that the mass of the accretion disk is $M_{disk} \ll M_{CS}$, 
the size of the accretion disk is given by (\ref{eq:Rdisk}). 
We found that the size of the accretion disk, does not depend on $V_{a}$ and is 
$R_{disk}  \approx 0.013 \; mas \approx 0.058 \; pc $.

\section{Circular orbit correction of C1 coordinates}
\label{COC_C1_1928+738}

We calculated the circular orbit correction for $M_{Cg} + M_{CS} = 10 * (M_{BHC6} + M_{BH4})$.

Using the parameters of the solution found in Sec.~\ref{Fit_C1_1928+738}, i.e. for 
$V_{a} = 0.1$ c, we have $M_{Cg} + M_{CS} \approx 1.9 \times 10^{9}$ $M_{\odot}$ and then 
$M_{BHC6} + M_{BH4} \approx 1.9 \times 10^{8}$ $M_{\odot}$. As the distance between the two BBH systems 
is $\approx 1.35$ $mas$ (see Sec.~\ref{Fit_C6_1928+738}), 
the corresponding orbital period of rotation of Cg-CS around BHC6-BH4 is $T_{bin} \approx 30320$ yr. 
Keeping the geometrical parameters of the solution found in Sec.~\ref{Fit_C1_1928+738} 
we calculate the trajectory and the tangent to the trajectory. At a given time, knowing 
the coordinates, $W_{CO}(t)$, $N_{CO}(t)$, of the trajectory of the VLBI component due 
to the slow circular orbit motion, and the coordinates, $W_{tan}(t)$, $N_{tan}(t)$, 
of the VLBI component along the tangent trajectory, the VLBI coordinates corrected from the slow 
orbital motion are given by equations (\ref{eq:W_cor}) and (\ref{eq:N_cor}).

\begin{figure}[ht]
\centerline{
\includegraphics[scale=0.5, width=8cm,height=6cm]{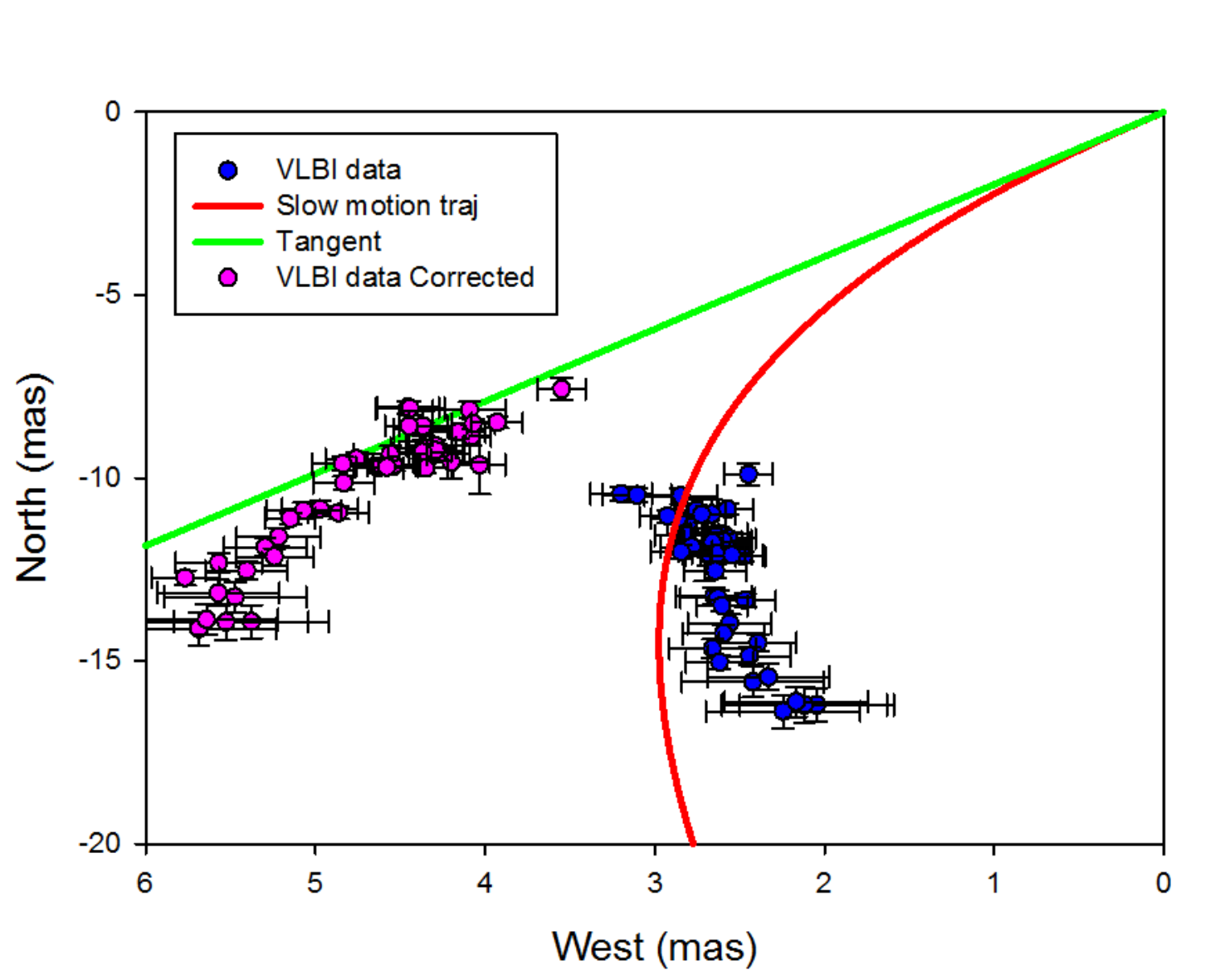}}
\caption{Plot of the trajectory of the VLBI component due to the slow circular orbit motion, 
the tangent trajectory, the VLBI coordinates given by \citet{KuGa+:14} and the coordinates 
corrected from the slow orbital motion.}
\label{fig:15_C1_Traj_15G_New4_COC_BH3_TpTb10_182_10_-+_Bis}
\end{figure}

We plotted in Fig.~\ref{fig:15_C1_Traj_15G_New4_COC_BH3_TpTb10_182_10_-+_Bis}, the trajectory of the 
VLBI component due to the slow circular orbit motion, the tangent trajectory, the VLBI coordinates 
given by \citet{KuGa+:14} and the coordinates corrected from the slow orbital motion.

Using the corrected VLBI coordinates, we made a new determination of the characteristics 
of the BBH system ejecting component C1. The result is given in Sec.~\ref{sec:solution_C1_1928+738}

\section{Fit of component C9}
\label{Fit_C9_1928+738}
We assumed that component C9 belongs to the family 
of components ejected by the black hole CS. 
To check this hypothesis and the consistency of the model found, we will use the characteristics 
of the BBH system Cg-CS and the characteristics of the geometrical parameters of the trajectory 
of C1, to fit the coordinates of components C9.

If C9 has been ejected by CS, we have to fit the coordinates of C9 using 
the characteristics of the BBH system Cg-CS found in Sec.~\ref{Fit_C1_1928+738}, i.e.
\begin{itemize}
  \item CS is the origin of the ejection,
	\item $T_{p} \approx 101866$ yr,
	\item $T_{p} / T_{b} \approx 53$,
	\item $R_{bin} \approx 0.220$ $\mu$as and 
	\item $M_{CS} /M_{Cg} \approx 0.25$,
\end{itemize}
and using the same geometrical parameters than those found to fit the trajectory of C1, i.e. 
\begin{itemize}
	\item $\Delta\Xi \approx 182^{\circ}$,
	\item $\Omega \approx 10.3^{\circ}$,
	\item $R_{o} \approx 56$ pc and  
	\item $T_{d}\approx 1278$ yr.
\end{itemize}

To begin, the coordinates of C9 given by \citep{KuGa+:14} are corrected by the offsets 
$\Delta X_{C9} \approx -0.07$ mas and $\Delta Y_{C9} \approx +0.21$ mas.

Then we calculate $\chi^{2}(i_{o})$ starting from $i_{o} \approx 23^{\circ}$ 
and assuming that the parameters:
\begin{itemize}
	\item $\phi_{o}$ the phase of the precession at $t_{o}$,
	\item $\gamma_{c}$ the bulk Lorentz,
	\item $\Psi_{o}$ the phase of the BBH system at $t_{o}$ and 
	\item $t_{o}$ the time origin of the ejection of the component,
\end{itemize}
are free parameters.

\begin{figure}[ht]
\centerline{
\includegraphics[scale=0.5, width=8cm,height=6cm]{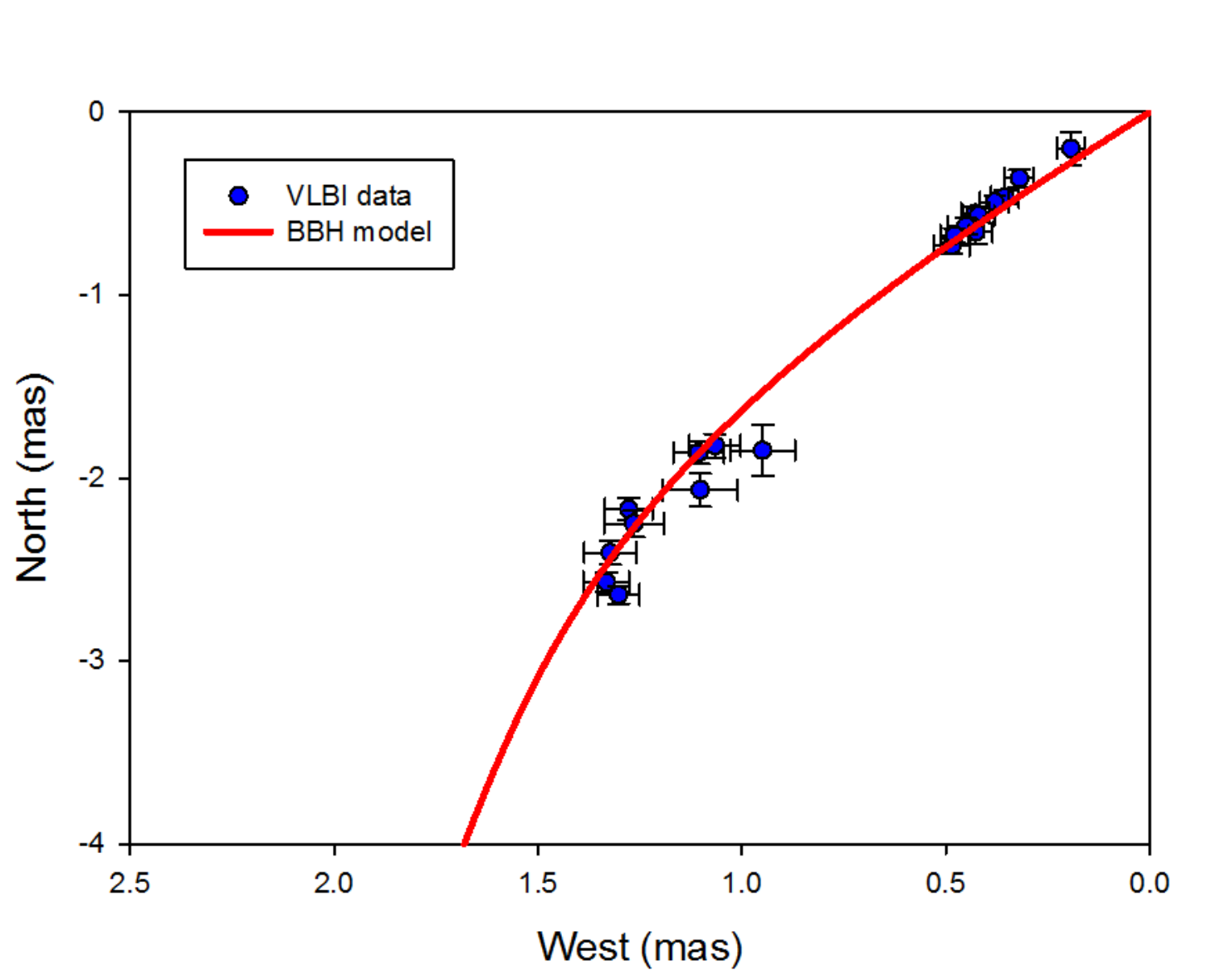}}
\caption{Trajectory of C9 assuming that it has been ejected by the black hole 
CS of the BBH system Cg-CS and using the characteristics of the BBH system Cg-CS obtained 
during the fit of component C1 and the geometrical parameters of the trajectory of C1.}
\label{fig:15a_comp9_v4_Traj_2_New_Off_CS_Short}
\end{figure}

The best fit is obtained for $i_{o} \approx 20^{\circ}$. The bulk Lorentz factor is 
$\gamma \approx 4.3$ and the time origin of the ejection is $t_{o} \approx 1993.3$. 
The trajectory of C9 is shown in Fig.~\ref{fig:15a_comp9_v4_Traj_2_New_Off_CS_Short}. 
We obtain a very good fit of each coordinate showing that 
\begin{itemize}
	\item component C9 has been ejected by CS, 
	\item the characteristics of the BBH system Cg-CS are correct and
	\item the solution found for the ejection of component C1 is the correct one.
\end{itemize}

\section{Fit of component C6}
\label{Fit_C6_1928+738}

\begin{figure}[ht]
\centerline{
\includegraphics[scale=0.5, width=8cm,height=12cm]{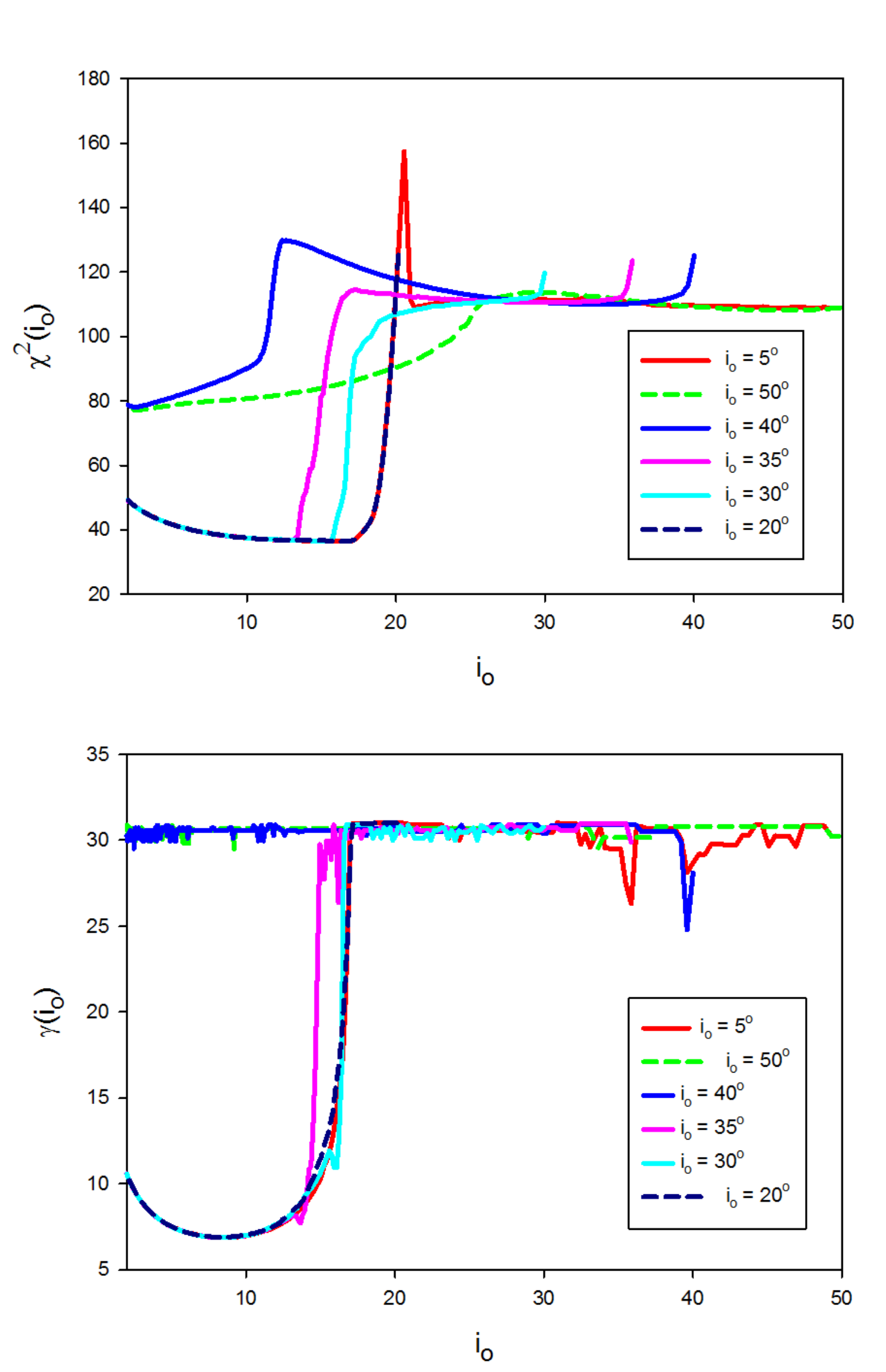}}
\caption{Assuming that component C6 is ejected by a single black hole, 
we applied the precession model and we calculated $\chi^{2}(i_{o})$ in the interval 
$2^{\circ} \leq i_{o} \leq 50^{\circ}$ starting from 6 different values of the inclination 
angle, namely $i_{o} = 5^{\circ}$, $50^{\circ}$,  $40^{\circ}$, $35^{\circ}$, $30^{\circ}$ 
and  $20^{\circ}$. We found that there exist solutions with $\gamma < 30$ only in the interval 
$2^{\circ} \leq i_{o} \leq 17^{\circ}$.
\textit{Top Figure}: The curves $\chi^{2}(i_{o})$ calculated starting from 6 different 
values of the inclination angle.
\textit{Bottom Figure}: The corresponding bulk Lorentz factor $\gamma$.}
\label{fig:16a_Chi2+Gamma_io_C6_Precession}
\end{figure}

\begin{figure}[ht]
\centerline{
\includegraphics[scale=0.5, width=8cm,height=12cm]{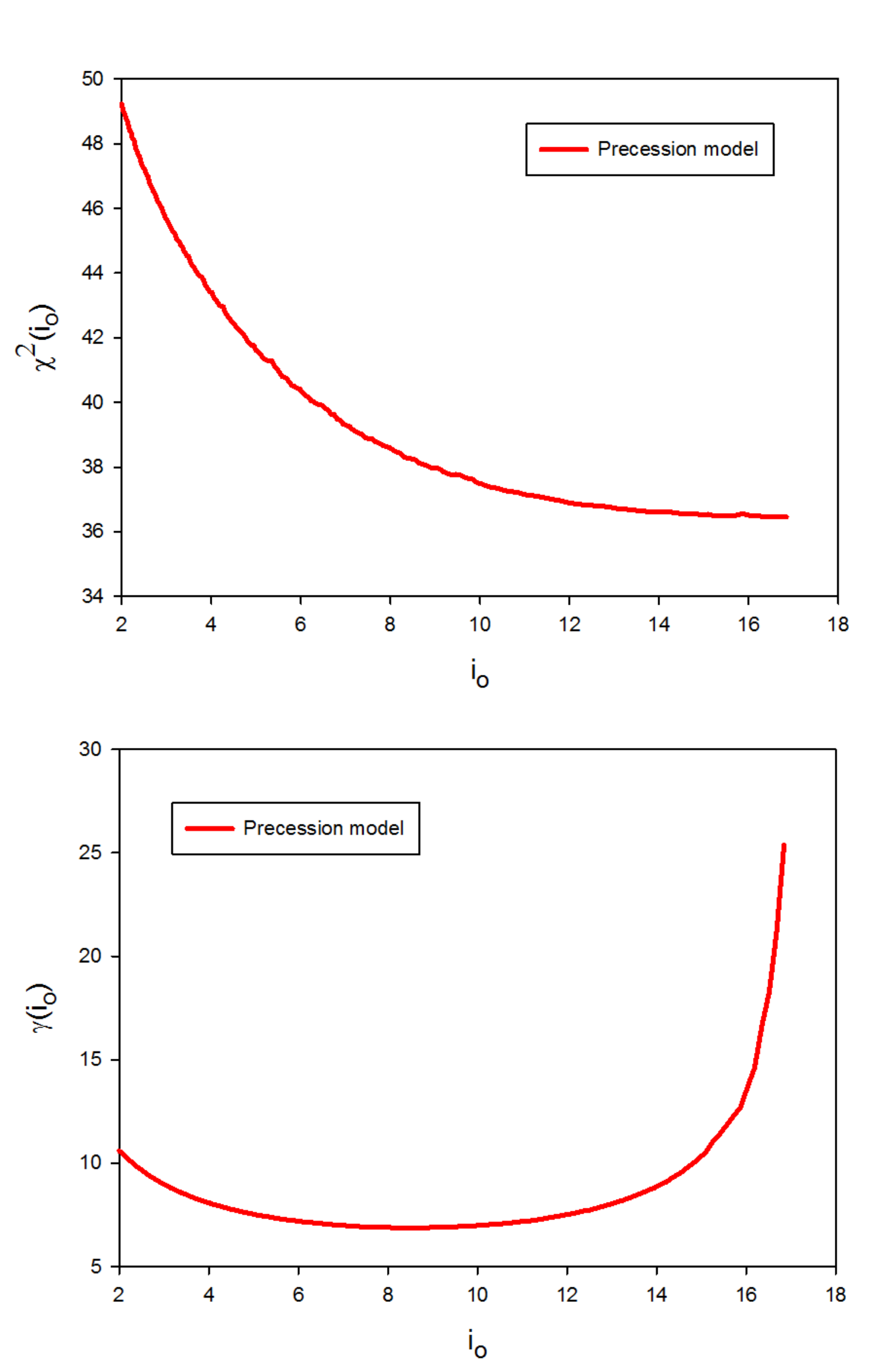}}
\caption{Assuming that component C6 is ejected by a single black hole, 
we applied the precession model and the solution with $\gamma < 30$ is a mirage solution, 
i.e. the curve $\chi^{2}(i_{o})$ is convex 	and it does not show a minimum; 
moreover the bulk Lorentz factor $\gamma$ diverges when 
$i_{o} \rightarrow 17^{\circ}$.
\textit{Top Figure}: The curve $\chi^{2}(i_{o})$ is convex 	and it does not show a minimum.
\textit{Bottom Figure}: The bulk Lorentz factor $\gamma$ diverges when 
	$i_{o} \rightarrow 17^{\circ}$.}
\label{fig:16_Chi2+Gamma_io_C6_Precession}
\end{figure}

The component C6 is not ejected by CS or Cg but is ejected by a third 
black hole. We can show that this third black hole belongs to a second BBH system. 
Indeed, if we assume that C6 is ejected by a single black hole, we applied the 
precession model and we studied the solution $\chi^{2}(i_{o})$ in the interval 
$2^{\circ} \leq i_{o} \leq 50^{\circ}$, we found that 
\begin{enumerate}
  \item there exist solutions with $\gamma < 30$ only in the interval 
	$2^{\circ} \leq i_{o} \leq 17^{\circ}$ (see Fig.~\ref{fig:16a_Chi2+Gamma_io_C6_Precession}), 
	\item the solution with $\gamma < 30$ is a mirage solution, 
	i.e. the curve $\chi^{2}(i_{o})$ is convex 	and it does not show a minimum; 
	moreover the bulk Lorentz factor $\gamma$ diverges when 
	$i_{o} \rightarrow 17^{\circ}$ (see Fig.~\ref{fig:16_Chi2+Gamma_io_C6_Precession}) and
	\item the precession period corresponding to the solution is: $1200$ yr $\leq T_{prec} \leq 2000$ yr. 
	This precession period too small to be explained by either the Lense-Thirring effect, 
	i.e. a spinning black hole or the precession due to the rotation of the black hole ejecting C6 
	around the BBH system Cg-CS.
\end{enumerate}

However, if the black hole ejecting C6 belongs to a second BBH system, the 
corresponding solution is no longer a mirage solution, i.e. the curve 
$\chi^{2}(i_{o})$ is concave and shows a minimum. We will call 
BHC6 the black hole ejecting component C6 and BH4 the second black hole of the 
second BBH system.\\

In this section we present the characteristics of the BBH system BHC6-BH4 
using the coordinates of C6 given by \citep{KuGa+:14}.

The main characteristics of the solution of the BBH system 
ejecting C6 are that
\begin{itemize}
    \item the coordinates of BHC6 are $X_{BHC6} \approx -0.11$ mas and $Y_{BHC6} \approx -1.30$ mas 
		(assuming that the origin is associated with Cg),
    \item none of the two black holes are associated with a stationary VLBI component, 
		i.e. they are not strong sources,
    \item the radius of the BBH system is $R_{bin} \approx 140$ $\mu as$ $\approx 0.62$ $pc$,
    \item calling $M_{BHC6}$ the mass of the black hole ejecting C6 and $M_{BH4}$ the mass of 
		the other black hole, the ratio $M_{BHC6}/M_{BH4}$ is $\approx 0.3$ , and
		\item the ratio $T_{p}/T_{b}$ is $\approx 1456$ .
\end{itemize}

We find that
\begin{itemize}
    \item the inclination angle is $i_{o} \approx 21^{\circ}$,
    \item the angle between the accretion disk and the rotation plane 
    of the BBH system is $\Omega \approx 3.6^{\circ}$,
    \item the bulk Lorentz factor of the VLBI component is $\gamma_{c} \approx 4.6$, and
    \item the time origin of the ejection of the VLBI component is $t_{o} \approx 1994.7$.
\end{itemize}

\begin{figure}[ht]
\centerline{
\includegraphics[scale=0.5, width=8cm,height=12cm]{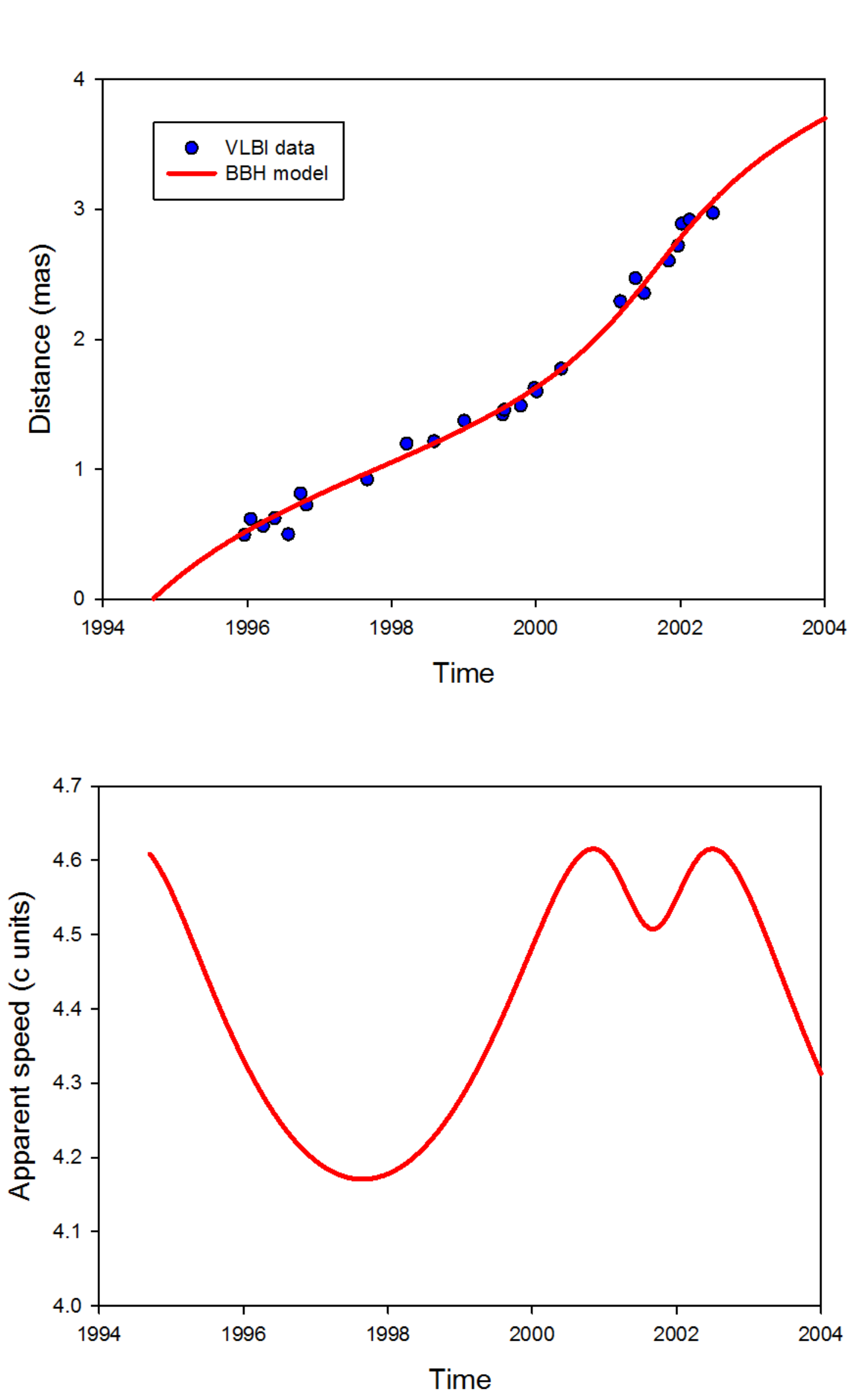}}
\caption{Variations of the distance and the apparent speed of component C6 
assuming a constant bulk Lorentz factor $\gamma_{c} \approx 4.6$. 
\textit{Top Figure}: From the plot of the variations of the distance we can deduce the mean speed: 
$3$ mas $/7.8$ yr $\approx 385$ $\mu$as /yr.
\textit{Bottom Figure}: Although the large value of the inclination angle, we 
observe superluminal motion with a mean speed $\approx 4.4$ c.}
\label{fig:16_Sol2_Comp6_Distance+Vap_Time_New1_BH3_Off}
\end{figure}

The variations of the distance and the apparent speed of component C6 are showed
in Fig.~\ref{fig:16_Sol2_Comp6_Distance+Vap_Time_New1_BH3_Off}. We find that component C6 moves with a 
mean apparent speed $v_{ap} \approx 4.4$ c, a value comparable to the one obtained by \citet{LiAl+:13}.

\begin{figure}[ht]
\centerline{
\includegraphics[scale=0.5, width=8cm,height=12cm]{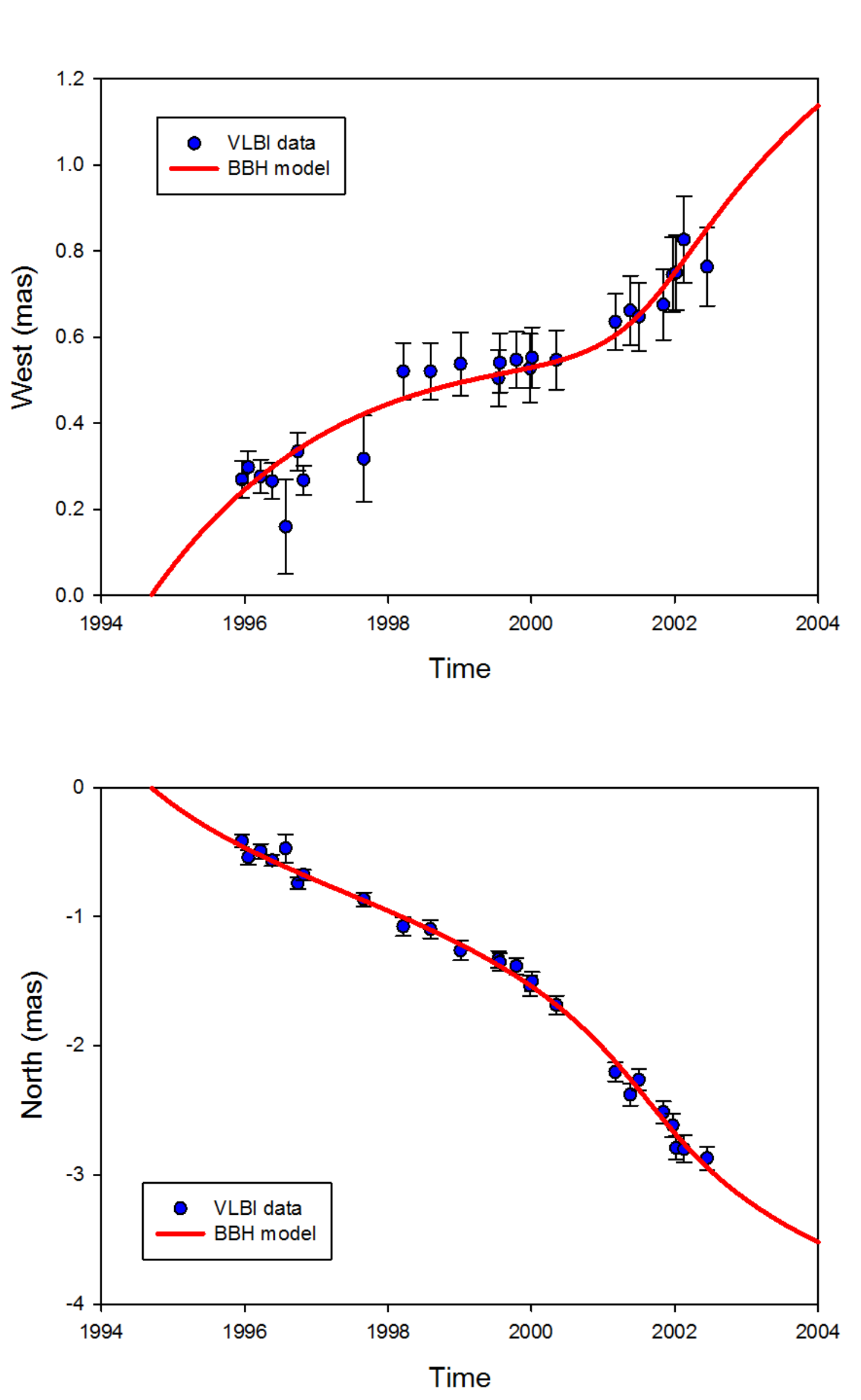}}
\caption{Fit of the two coordinates $W(t)$ and $N(t)$ of component C6 of 1928+738. 
They correspond to the solution with $T_{p}/T_{b} \approx 1500$, 
$M_{BHC6}/ M_{BH4} \approx 0.3$, and $i_{o} \approx 21^{\circ}$. The points are the observed 
coordinates of component C6 that were corrected by the offsets $\Delta W \approx +110$ $\mu as$ 
and $\Delta N \approx +1300$ $\mu as$. The VLBI coordinates and their error bars 
are taken from \citet{KuGa+:14}. The red lines are the coordinates of the 
component trajectory calculated using the BBH model.}
\label{fig:17_Sol2_Comp6_Xt+Yt_New1_BH3_Off}
\end{figure}

The fit of the two coordinates $W(t)$ and $N(t)$ of the component C6 of 1928+738 is  
showed in Fig.~\ref{fig:17_Sol2_Comp6_Xt+Yt_New1_BH3_Off}. 
The points are the observed coordinates of component C6 that were corrected 
by the offsets $\Delta W \approx +110$ $\mu as$ and $\Delta N \approx +1300$ $\mu as$, and 
the red lines are the coordinates of the component trajectory calculated using the BBH model.

\subsection{Determining the family of solutions}
For the inclination angle previously found, i.e., 
$i_{o}\approx 21^{\circ}$, $T_{p}/T_{b} \approx 1500$, $M_{BHC6}/M_{BH4} \approx 0.3$, 
and $R_{bin} \approx 140$ $\mu as$, we gradually varied $V_{a}$ between $0.001$ c 
and $0.45$ c. The function $\chi^{2}(V_{a})$ remained constant,  
indicating a degeneracy of the solution. We deduced the 
range of variation of the BBH system parameters. They are given in Table 6.

\begin{center}
Table 6 : Ranges for the BBH system parameters ejecting C6\medskip%

\begin{tabular}
[c]{c||c|c}\hline
$V_{a}$                    & $0.001 \: c$                          & $0.45 \: c$                            \\\hline
$T_{p}(V_{a})$             & $\approx 149000000$ yr                & $\approx 179500 $ yr                    \\\hline
$T_{b}(V_{a})$             & $\approx 102500$ yr                   & $\approx 123$ yr                       \\\hline
$(M_{Cg}+ M_{CS})(V_{a})$  & $\approx 2 \; 10^{5}$ $M_{\odot}$     & $\approx 1.4 \; 10^{11}$ $M_{\odot}$   \\\hline
\end{tabular}
\end{center}

Table 6 provides the range of the BBH system parameters ejecting C6. To obtain the final range of the 
two BBH systems Cg-CS and BHC6-BH4 one has to make the intersection of the ranges of BBH systems parameters 
found after the fits of C8, C1 and C6 (see Sec.~\ref{section:Discussion_conclusion}).

For $V_{a,CS} = 0.1$ c, we find that the total mass of the BBH system ejecting C1 is 
$M_{Cg} + M_{CS} \approx 1.87 \times 10^{9}$ $M_{\odot}$. The total mass of the BBH system 
ejecting C6 is $M_{BHC6} + M_{BH4} = (M_{Cg}+M_{CS})/10$ if $V_{a,BHC6} \approx 0.030$ c, 
i.e. the propagation speeds of the perturbations are different for different families of trajectories 
(see Sec.~\ref{section:Discussion_conclusion} for the determination of the propagation speeds of the 
perturbations of three families of trajectories if $M_{Cg} + M_{CS} \approx 8 \times 10^{8}$ $M_{\odot}$).

\subsection{Determining the size of the accretion disk}

From the knowledge of the mass ratio $M_{BHC6}/M_{BH4} \approx 0.3$ and 
the ratio $T_{p}/T_{b} \approx 1500$, we calculated in the previous section 
the mass of the ejecting black hole $M_{BHC6}$, the orbital period $T_{b}$, 
and the precession period $T_{p}$ for each value of $V_{a}$.

The rotation period of the accretion disk, $T_{disk}$, is given by 
(\ref{eq:Tdisk}). Thus we calculated the rotation period of the accretion disk, and 
assuming that the mass of the accretion disk is $M_{disk} \ll M_{BHC6}$, 
the size of the accretion disk is given by (\ref{eq:Rdisk}). 
We found that the size of the accretion disk, does not depend on $V_{a}$ and is 
$R_{disk}  \approx 0.00096 \; mas \approx 0.0043 \; pc $.

\section{Circular orbit correction of C6 coordinates}
\label{COC_C6_1928+738}

We calculated the circular orbit correction for $M_{BHC6} + M_{BH4} = (M_{Cg}+M_{CS})/10$.

Using the parameters of the solution found in Sec.~\ref{Fit_C6_1928+738}, i.e. for   
$V_{a,C6} = 0.1$ c, the mass of the BBH system ejecting C6 is 
$M_{BHC6}+M_{BH4} \approx 2.5 \times 10^{9}$ $M_{\odot}$ and then 
$M_{Cg} + M_{CS} \approx 2.5 \times 10^{10}$ $M_{\odot}$. 
We could calculate the circular orbit correction for a different 
value of $V_{a}$ to have a mass $M_{Cg} + M_{CS}$ equal to the mass used in Sec.~\ref{COC_C8_1928+738}, 
however due to the degeneracy of the solution the result will be the same.

The distance between the two BBH systems is $\approx 1.35$ $mas$ (see Sec.~\ref{Fit_C6_1928+738}), 
the corresponding orbital period of rotation of Cg-CS around BHC6-BH4 is $T_{bin} \approx 8837$ yr. 
Keeping the geometrical parameters of the solution found in Sec.~\ref{Fit_C6_1928+738} 
we calculate the trajectory and the tangent to the trajectory. At a given time, knowing 
the coordinates, $W_{CO}(t)$, $N_{CO}(t)$, of the trajectory of the VLBI component due 
to the slow circular orbit motion, and the coordinates, $W_{tan}(t)$, $N_{tan}(t)$, 
of the VLBI component along the tangent trajectory, the VLBI coordinates corrected from the slow 
orbital motion are given by equations (\ref{eq:W_cor}) and (\ref{eq:N_cor}).

\begin{figure}[ht]
\centerline{
\includegraphics[scale=0.5, width=8cm,height=6cm]{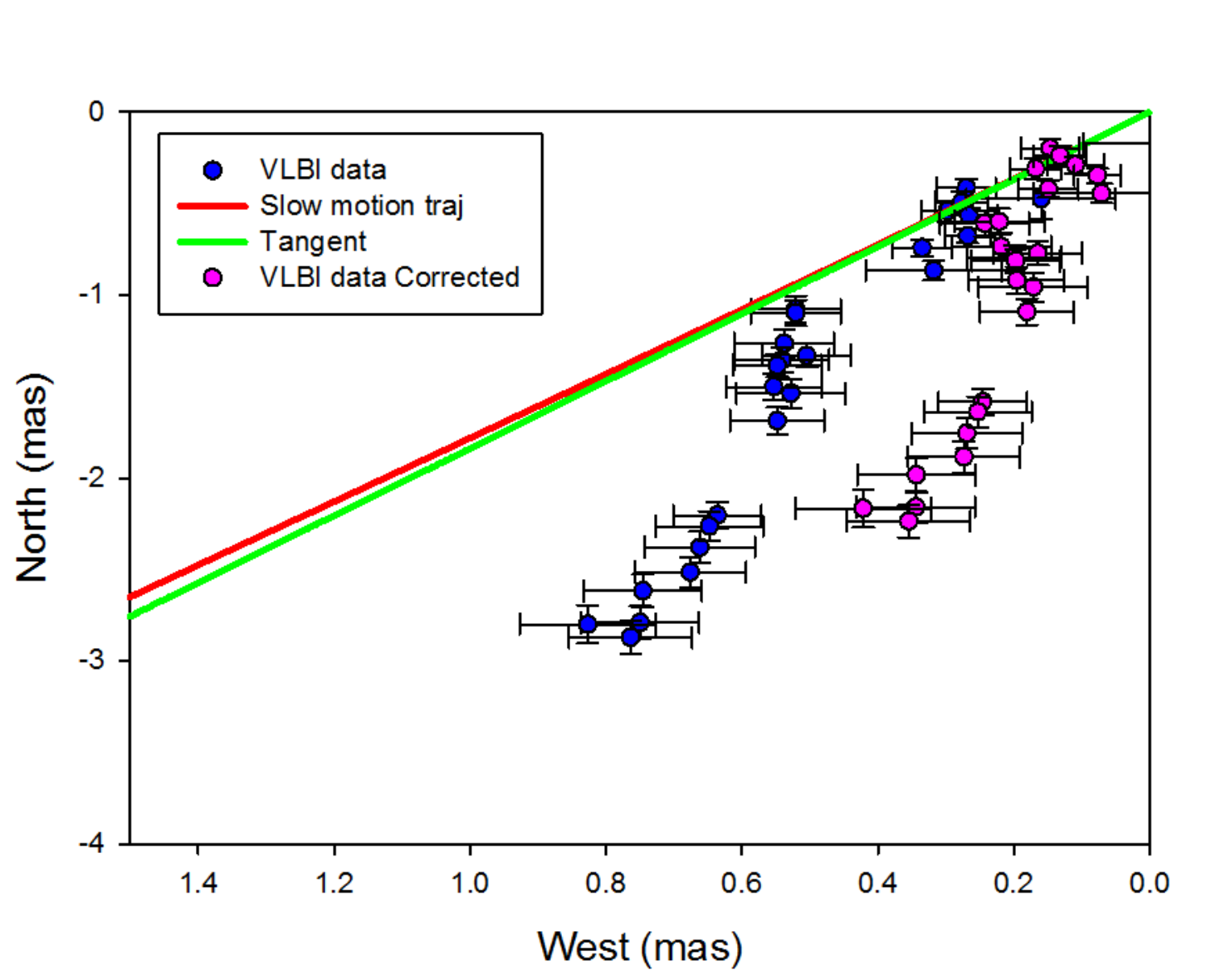}}
\caption{Plot of the trajectory of the VLBI component due to the slow circular orbit motion, 
the tangent trajectory, the VLBI coordinates given by \citet{KuGa+:14} and the coordinates 
corrected from the slow orbital motion.}
\label{fig:18_C6_Traj_15G_BH3_COC_165_01_-+_Bis}
\end{figure}

We plotted in Fig.~\ref{fig:18_C6_Traj_15G_BH3_COC_165_01_-+_Bis}, the trajectory of the 
VLBI component due to the slow circular orbit motion, the tangent trajectory, the VLBI coordinates 
given by \citet{KuGa+:14} and the coordinates corrected from the slow orbital motion.

Using the corrected VLBI coordinates, we made a new determination of the characteristics 
of the BBH system ejecting component C6. The result is given in Sec.~\ref{sec:solution_C6_1928+738}.

\section{Fit of component C5}
\label{Fit_C5_1928+738}
We assumed that component C5 belongs to the family 
of components ejected by the black hole BHC6. 
To check this hypothesis and the consistency of the model found, we will use the characteristics 
of the BBH system BHC6-BH4 and the characteristics of the geometrical parameters of the trajectory 
of C6, to fit the coordinates of components C5.

If C5 has been ejected by BHC6, we have to fit the coordinates of C5 using 
the characteristics of the BBH system BHC6-BH4 found in Sec.~\ref{Fit_C6_1928+738}, i.e.
\begin{itemize}
  \item BHC6 is the origin of the ejection,
	\item $T_{p} \approx 1344545$ yr,
	\item $T_{p} / T_{b} \approx 1456$,
	\item $R_{bin} \approx 0.140$ $\mu$as and 
	\item $M_{BHC6} /M_{BH4} \approx 0.3$,
\end{itemize}
and using the same geometrical parameters than those found to fit the trajectory of C6, i.e. 
\begin{itemize}
	\item $\Delta\Xi \approx 165^{\circ}$,
	\item $\Omega \approx 3.6^{\circ}$,
	\item $R_{o} \approx 103$ pc and  
	\item $T_{d}\approx 1500$ yr.
\end{itemize}

To begin, the coordinates of C5 given by \citep{KuGa+:14} are corrected by the offsets 
$\Delta X_{C5} \approx +0.10$ mas and $\Delta Y_{C5} \approx +1.30$ mas.

Then we calculate $\chi^{2}(i_{o})$ starting from $i_{o} \approx 21^{\circ}$ 
and assuming that the parameters:
\begin{itemize}
	\item $\phi_{o}$ the phase of the precession at $t_{o}$,
	\item $\gamma_{c}$ the bulk Lorentz,
	\item $\Psi_{o}$ the phase of the BBH system at $t_{o}$ and 
	\item $t_{o}$ the time origin of the ejection of the component,
\end{itemize}
are free parameters.

\begin{figure}[ht]
\centerline{
\includegraphics[scale=0.5, width=8cm,height=6cm]{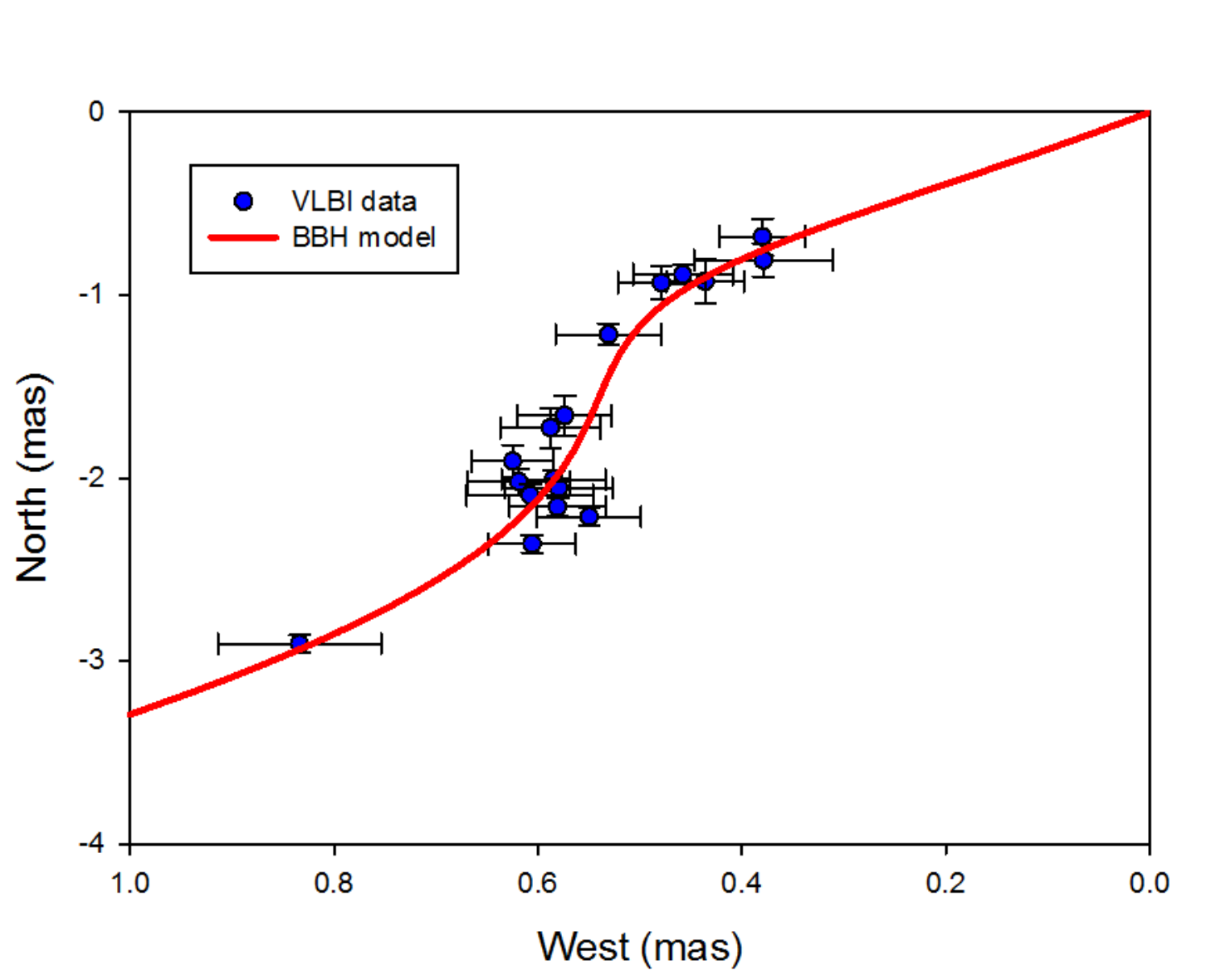}}
\caption{Trajectory of C5 assuming that it has been ejected by the black hole 
BHC6 of the BBH system BHC6-BH4 and using the characteristics of the BBH system BHC6-BH4 obtained 
during the fit of component C6 and the geometrical parameters of the trajectory of C6.}
\label{fig:19_comp5_v4_2_Traj_New}
\end{figure}

The best fit is obtained for $i_{o} \approx 20^{\circ}$. The bulk Lorentz factor is 
$\gamma \approx 4.3$ and the time origin of the ejection is $t_{o} \approx 1991$. 
The trajectory of C5 is shown in Fig.~\ref{fig:19_comp5_v4_2_Traj_New}. 
We obtain a very good fit of each coordinate showing that 
\begin{itemize}
	\item component C5 has been ejected by BHC6, 
	\item the characteristics of the BBH system BHC6-BH4 are correct and
	\item the solution found for the ejection of component C6 is the correct one.
\end{itemize}

\end{document}